\documentclass[acmlarge]{acmart}
\usepackage{balance}
\usepackage{booktabs} 

\usepackage[T1]{fontenc}
\usepackage[ruled]{algorithm2e}
\SetAlFnt{\algofont}
\SetAlCapFnt{\algofont}
\SetAlCapNameFnt{\algofont}
\SetAlCapHSkip{0pt}
\IncMargin{\parindent}

\usepackage{amssymb}
\usepackage{graphicx}
\usepackage{tikz}
\usetikzlibrary{arrows,calc,positioning}
\usetikzlibrary{automata}
\usepackage{pgfplots}
\pgfplotsset{compat=newest}
\usetikzlibrary{plotmarks}
\usepackage{grffile}
\usepackage{amsmath,amssymb}
\usetikzlibrary{shapes.gates.logic.US,shapes.gates.logic.IEC,calc}
\usepackage{url}
\usepackage{pdflscape}
\usepackage{afterpage}
\usepackage{psfrag}
\usepackage{epstopdf}
\usepackage{caption}
\DeclareCaptionType{copyrightbox}
\usepackage{ifthen}
\usepackage{cleveref}
\usetikzlibrary{pgfplots.groupplots}

\usepackage{booktabs} 
\usepackage[ruled]{algorithm2e} 

\SetAlFnt{\small}
\SetAlCapFnt{\small}
\SetAlCapNameFnt{\small}
\SetAlCapHSkip{0pt}
\IncMargin{-\parindent}

\newcommand{\po}{\mathrm{P}}     
\newcommand{\ro}{\mathrm{R}}     
\newcommand{\co}{\mathrm{C}}     
\newcommand{\fo}{\mathrm{F}}     
\newcommand{\so}{\mathrm{S}}     
\newcommand{\Rl}{\mathbb{R}}     
\newcommand{\Rt}{\textbf{R}}	 
\newcommand{\Na}{\mathbb{N}}     
\newcommand{\F}{$\mathcal{F}$}   
\newcommand{\Tcln}{T_{cln}}      
\newcommand{\Trpl}{T_{rplc}}     
\newcommand{\Td}{T_{deg}}        
\newcommand{\Toh}{T_{oh}}        
\newcommand{\Tin}{T_{insp}}        
\newcommand{\Trp}{T_{rep}}       
\newcommand{\Act}{\text{TL}}    
\newcommand{\Ap}{\text{AP}}		 
\newtheorem{remark}{Remark}
\newtheorem*{example*}{Example}
\pgfplotsset{every tick label/.append style={font=}}

\acmJournal{TOSN}
\acmVolume{1}
\acmNumber{1}
\acmArticle{1}
\acmYear{2018}
\acmMonth{1}
\acmArticleSeq{1}

\definecolor{mycolor1}{rgb}{0.00000,0.44700,0.74100}%
\definecolor{mycolor2}{rgb}{0.85000,0.32500,0.09800}%
\definecolor{mycolor3}{rgb}{0.92900,0.69400,0.12500}%
\definecolor{mycolor4}{rgb}{0.49400,0.18400,0.55600}%
\definecolor{mycolor5}{rgb}{0.46600,0.67400,0.18800}%
\definecolor{mycolor6}{rgb}{0.30100,0.74500,0.93300}%
\definecolor{mycolor7}{rgb}{0.63500,0.07800,0.18400}%

\begin{document}
	\title{Maintenance of Smart Buildings using Fault Trees}

	\author[first]{Nathalie Cauchi}

	\affiliation[first]{%
		\institution{Department of Computer Science, University of Oxford}
		\streetaddress{Parks Rd}
		\city{Oxford}}
	\email{name.surname@cs.ox.ac.uk}
	\author[first]{Khaza Anuarul Hoque}
	\affiliation{
			\institution{Department of Computer Science, University of Oxford}
			\streetaddress{Parks Rd}
			\city{Oxford}}
	\author{Marielle Stoelinga} 
	\affiliation{%
		\institution{{Formal Methods and Tools Group}, University of Twente, The Netherlands; Department of Software Science, Radboud University, The Netherlands }}
	\email{marielle@cs.utwente.nl}	
	\author[first]{Alessandro Abate }
		\affiliation{
		\institution{Department of Computer Science, University of Oxford}
		\streetaddress{Parks Rd}
		\city{Oxford}}

	\begin{abstract}
		Timely maintenance is an important means of increasing system dependability and life span.  
		Fault Maintenance trees (FMTs) are an innovative framework incorporating both maintenance strategies and degradation models and serve as a good planning platform for balancing total costs (operational and maintenance) with dependability of a system. In this work, we apply the FMT formalism to a {Smart Building} application and propose a framework that efficiently encodes the FMT into Continuous Time Markov Chains. This allows us to obtain system dependability metrics such as system reliability and mean time to failure, as well as costs of maintenance and failures over time, for different maintenance policies. We illustrate the pertinence of our approach by evaluating various dependability metrics and maintenance strategies of a Heating, Ventilation and Air-Conditioning system.	\footnote{Parts of this paper have been published in the 4th ACM International Conference on Systems of Energy-Efficient Build Environments (BuildSys 2017)~\cite{cauchi2017efficient}.}
	
	\end{abstract}



	\maketitle

	\renewcommand{\shortauthors}{N.Cauchi et al.}
	\section{Introduction}
	\label{sec:Intro}
	{The Internet-of-things has enabled a new type of buildings, termed \textit{Smart Buildings}, which aim to deliver useful building services that are cost effective, reliable, ubiquitous and ensure occupant comfort and productivity (thermal quality, air comfort). Smart buildings are equipped with many sensors such that a high level of intelligence is achieved: light and heating can be switched on automatically; fire and burglar alarms can be more sophisticated; and cleaning services can be connected to the occupancy rate. Maintenance is a key element to keep smart buildings smart: 
    without proper maintenance (cleaning, replacements, etc.), the benefits of achieving greater efficiency, comfort, increased building lifespan, reliability and sustainability are quickly lost. 
	
	In this paper, we consider an important element in smart buildings, namely the heating, ventilation and air-conditioning (HVAC) system, responsible for maintaining thermal comfort and ensuring good air-quality in buildings. }
	One way of improving the lifespan and reliability of such systems is by employing methods
	to detect faults and 
	to perform preventive and predictive maintenance actions. 
	Techniques for fault detection and diagnosis for Smart Building applications have been developed in
	~\cite{Yan2017,boem2017distributed}. Predictive and preventive maintenance strategies are devised in~\cite{cauchi2017model,macek2017long,Babishin201610480}. 
	Moreover, a reliability-centered predictive maintenance policy is proposed in~\cite{zhou2007reliability}. This policy is for a continuously monitored system which is subject to degradation due to imperfect maintenance.
	However, these techniques neglect reliability measurements and focus only on synthesis of maintenance policies in the presence of degradation and faults.
	{The current industrial standard for measuring a system's  reliability is the use of Fault trees, where the focus is on finding the root causes of a system failure using a top-down approach and do not incorporate degradation of system components and maintenance action~\cite{ammar2016formal,ruijters2015fault,Li20151400}. }
    \cite{ruijters2016fault} presents the Fault Maintenance Tree (FMT) as a framework that allows to perform planning strategies for balancing total costs and reliability and availability of the system. FMTs are 
	an extension of FT encompassing both degradation and maintenance models. The degradation models represent the different levels of component degradation and are known as Extended Basic Events (EBE). The maintenance models incorporate the undertaken maintenance policy which includes both inspections and repairs. These are modelled using Repair and Inspection modules in the FMT framework. 
	
	{In literature, analysis of FMTs is performed using Statistical Model checking (SMC)~\cite{ruijters2016fault}, which 
	generates sample executions of a stochastic system according to the distribution defined
	by  the  system and computes statistical guarantees based on the executions~\cite{legay2010statistical}.} In contrast, Probabilistic Model Checking (PMC) provides formal guarantees with higher accuracy when compared with SMC~\cite{younes2006numerical}, at a cost of being 
	more memory intensive and may result in a state space  explosion. 
	PMC is an automatic procedure for establishing if a
	desired property holds in a probabilistic system model which encodes the probability of making a transition between states. This allows for making quantitative
	statements about the system's behaviour which are expressed as probabilities
	or expectations~\cite{kwiatkowskaadvances}. 
Probabilistic model checking has been successfully applied in a different
domains so far including:{ aerospace and avionics~\cite{hoque2017formal},
optical communication~\cite{siddique2017formal}, systems biology~\cite{dannenberg2013dna} and robotics~\cite{feng2015controller}}. In this paper we tackle the FMT analysis using PMC. 
	Our contributions can be summarised as follows: 
	\begin{enumerate}
		\item We formalise the FMT using Continuous Time Markov Chain (CTMCs) 
		 and the dependability metrics of a Heating, Ventilation and Air-Conditioning (HVAC) system,  
		using the Continuous Stochastic Logic (CSL) formalism, such that they can be computed using the PRISM model checker~\cite{PRISM}.
		\item To tackle the state space explosion problem, we present an FMT abstraction technique which decomposes a large FMT into an equivalent abstract  FMT based on a graph decomposition algorithm. This involves an intermediate step where the large FMT is transformed into an equivalent direct acyclic graph and decomposed into a set of small sub-graphs. Each of these small sub-graphs are converted to an equivalent smaller CTMC and analysed separately to compute the required metric, while maintaining the original FMT hierarchy. 
	 Using our framework, we are able to achieve a 67$\%$ reduction in the state space  size.
		\item Finally, we construct a FMT that identifies failure of an HVAC and illustrate the use of the developed framework to construct and analyse the FMT. We also evaluate relevant performance metrics using the PRISM model checker, compare different maintenance strategies and highlight the importance of performing maintenance actions. 
	\end{enumerate}
	

	{This article has the following structure:  Section~\ref{sec:hvac} introduces the heating, ventilation and air-conditioning (HVAC) set-up under consideration together with the maintenance question we are addressing. This is followed by Section~\ref{sec:Prelim}, which presents the fault maintenance trees and probabilistic model checking frameworks. Next, we present the
  developed methodology for modelling FMT using CTMCs {and perform model checking in Section~\ref{subsec:FMTFrame:Modelling}}. 
	The framework is then applied to the HVAC system in Section~\ref{sec:CaseStudy}.}
	
	\section{Problem formulation}
	\label{sec:hvac}
\begin{figure*}[ht!]
		\centering
		\includegraphics[width=.75\textwidth]{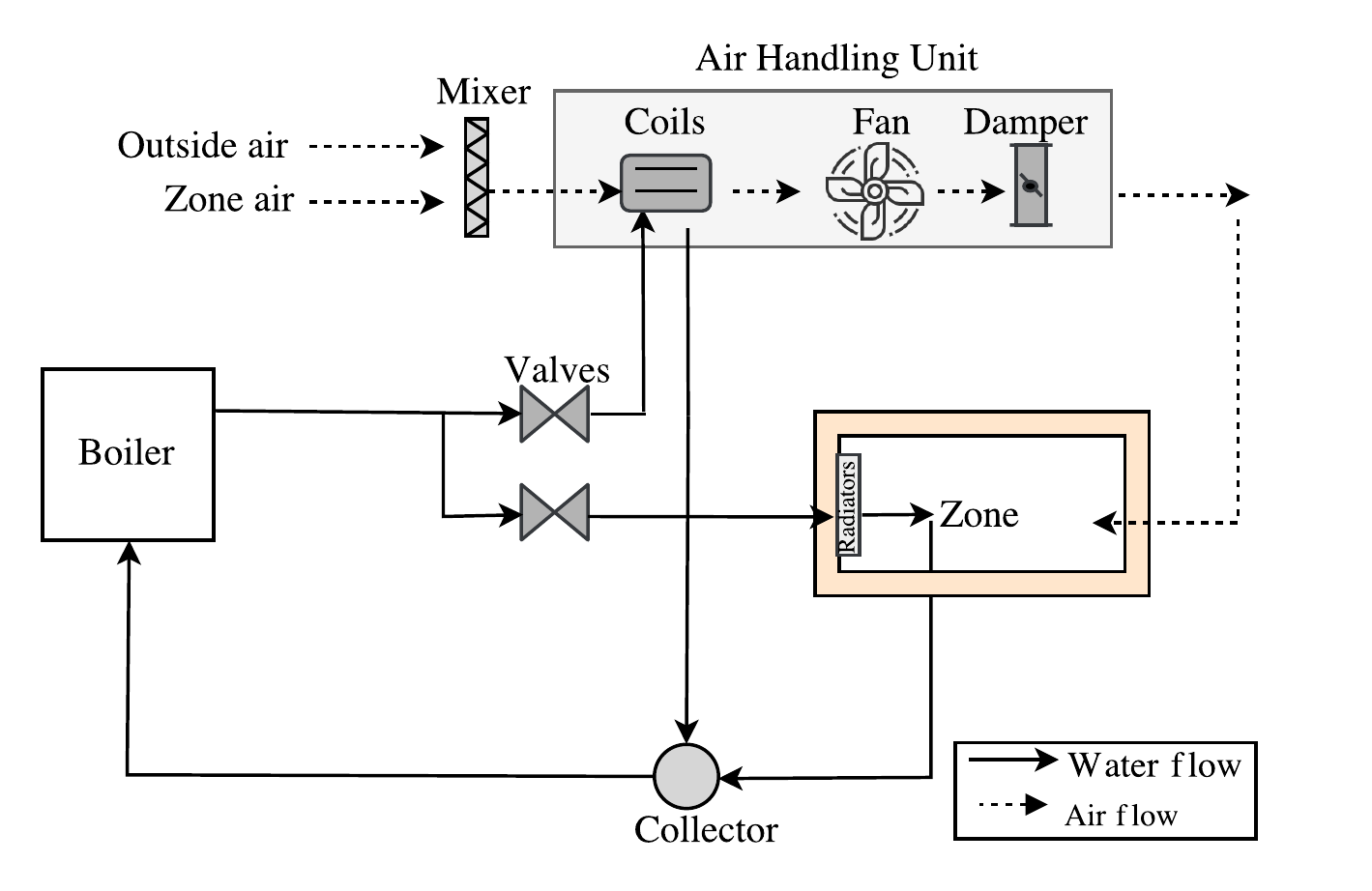}
		\caption{{High level schematic of an  HVAC system. }}
		\label{fig:hvac-system}
	\end{figure*}

	We consider the heating, ventilation and air-conditioning (HVAC) system setup 
	found within the Department of Computer Science, at the University of Oxford.
	A graphical description is shown in Figure~\ref{fig:hvac-system}. It is composed of two circuits - the air flow circuitry and the water circuit. The gas boiler heats up the supply water and transfers the supply water into two sections - the supply air heating coils and the radiators. 
	The rate of water flowing in the heating coil is controlled using a heating coil valve, while the rate of water flow in the radiator is controlled using a separate valve. {The outside air is mixed with the air extracted from the zone via the mixer}. This is fed into the heating coil, which warms up the input air to the desired supply air temperature. This air is supplied back, at a rate controlled by the Air Handling unit (AHU) dampers, into the zone via the supply fan.  The radiators are directly connected to the water circuitry and transfer the heat from the water into the zone. The return water, from both the heating coils and the radiators, {is then  passed through the collector } and is returned back to the boiler. 
	
	The correct maintenance of this system is essential to ensure that the building operates with optimum efficiency while user comfort is maintained. The choice on the type of maintenance depends on several factors, including the different costs of maintenance and failures, and the {practical feasibility} of performing maintenance. To this end, we aim to address the following maintenance questions: (1) what is the optimal maintenance strategy that {minimises system failures?}; (2) what is the best trade-off between cost of inspections, operation and maintenance, vs the system's number of expected failures?; (3) how frequently should the different maintenance actions such as performing a cleaning or a replacement be performed?; and (4) what is the effect of employing maintenance over a specific time horizon vs not performing maintenance?

	\section{Preliminaries}
	\label{sec:Prelim}

	\subsection{Fault trees}
	\label{subsec:Prelim:FT}
	{Fault trees (FT) are directed acyclic graphs (DAG) describing the combinations of component failures that lead to system failures. It consists of two types of nodes: events and gates. 
	\begin{definition}[Event]
		\label{def:event}
		An {event} is
		an occurrence within the system, typically the failure of a subsystem down to an individual component.  Events can be  divided into	\textit{basic events} (BEs) and intermediate events. BE occur spontaneously and denote the component/system failures while intermediate events are caused by one of or more other events. The  event at  the  top  of  the  tree, called  the \textit{top  event}  (TE), is  the  event  being  analysed, modeling the failure of the (sub)system under consideration {(both type of events are highlighted in Figure~\ref{fig:FT})}. 
	\end{definition}
	\begin{definition}[Gates]
	\label{def:gates}
	The internal nodes of the graph are called \textit{gates} and describe the different ways that failures can interact to cause other components to fail i.e.   how  failures  in  subsystems  can  combine  to cause a system failure.  Each gate has one output and one or more inputs.  The gates in a FT can be of several types and these include the AND gate, OR gate, k/N-gate~\cite{ruijters2016fault}. The output of a gate triggers an event which shows how failures propagate through the fault tree. 
\end{definition}
Figure~\ref{fig:FT} depicts a fault tree were the basic events are shown using circles, top and intermediate events are depicted  by  a  rectangle.
	\begin{figure}[h!]
		\centering
		\includegraphics[width=0.55\columnwidth]{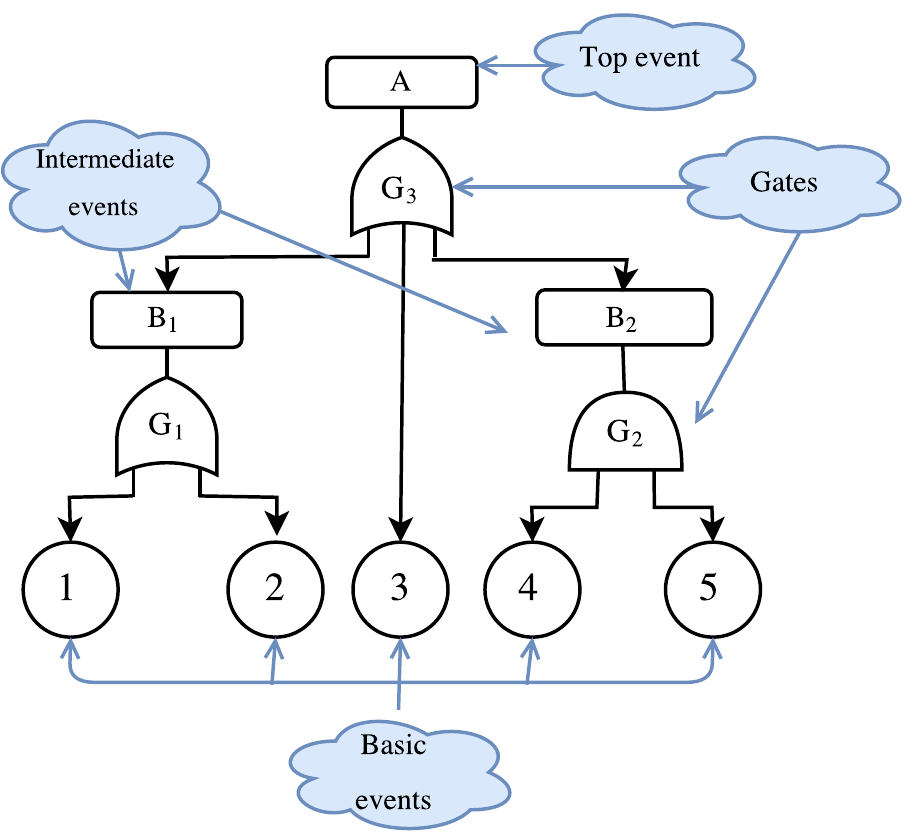}
		\caption{Example of a FT with five basic events (1-5), two intermediate events ($B_1, B_2$) and top event $A$; failures are
			propagated by the gates ($G_1 - G_3$). }
		\label{fig:FT}
	\end{figure}}
	\subsection{Fault maintenance trees}
	\label{subsec:Prelim:FMTFrame}
\begin{figure}[h!]
		\centering
		\includegraphics[width=0.8\columnwidth]{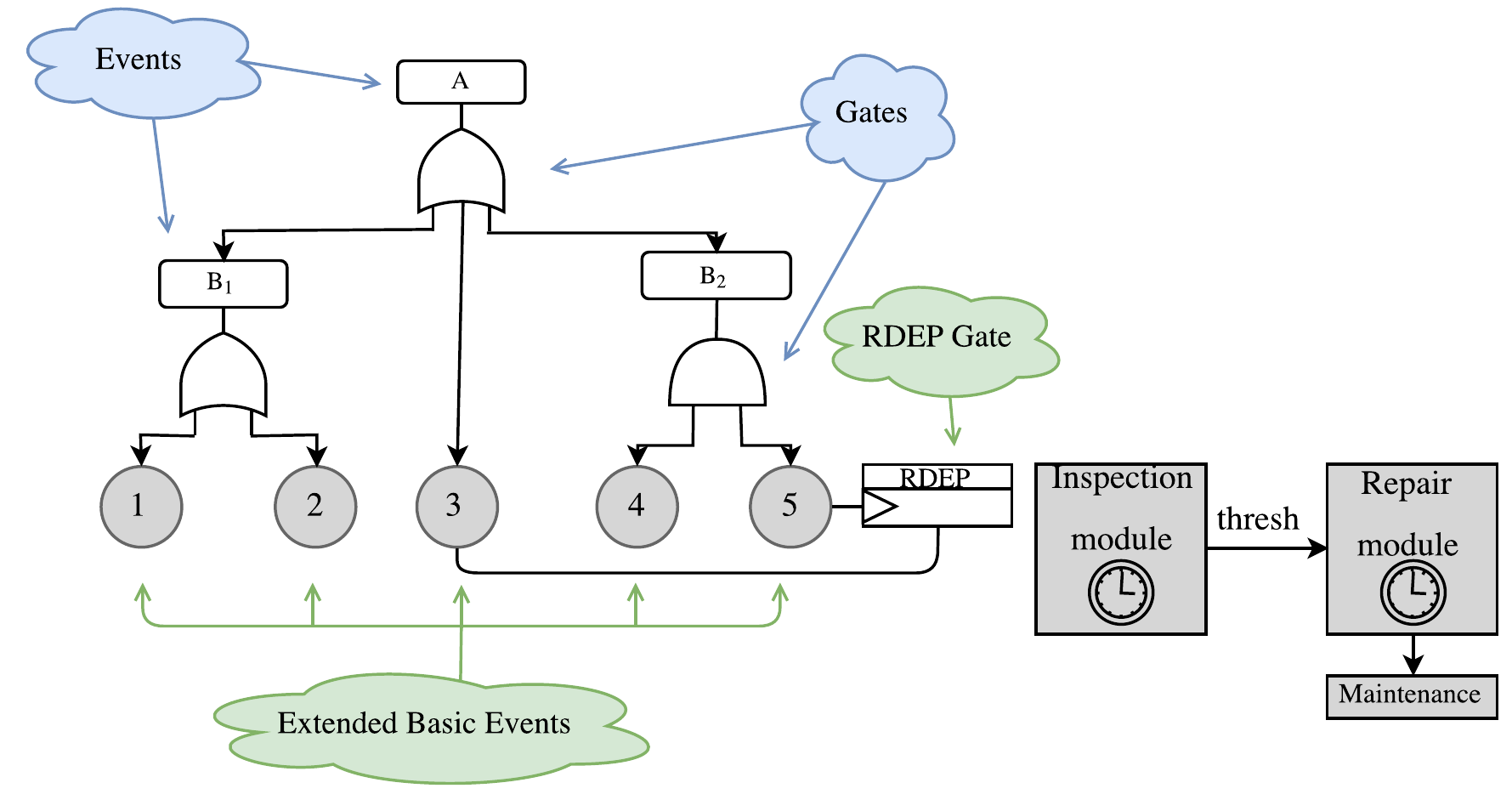}
		\caption{Example of a fault maintenance tree.}
		\label{fig:FMTsmall}
	\end{figure}

	Fault maintenance trees (FMT) extend fault trees by including maintenance (all the standard FT gates are also employed by the FMTs). This is achieved by making use of: 
	\begin{enumerate}
		\item \textit{Extended Basic Events (EBE)} - The basic events are modified to incorporate degradation models of the component the EBE represents. The degradation models represent different discrete levels of degradations the components can be in and are a function of time. The timing diagram showing the progression of degradation within an EBE is shown in Figure~\ref{fig:EBETiming}. The presented EBE had $N$ discrete degradation levels, initially the EBE is its new state and it gradually moves from one degradation levels, based on the underlying distribution describing the degradation, to the next until the faulty level {$N$} is reached. 
		\begin{figure}[h!]
			\centering
			\includegraphics[width=0.35\columnwidth]{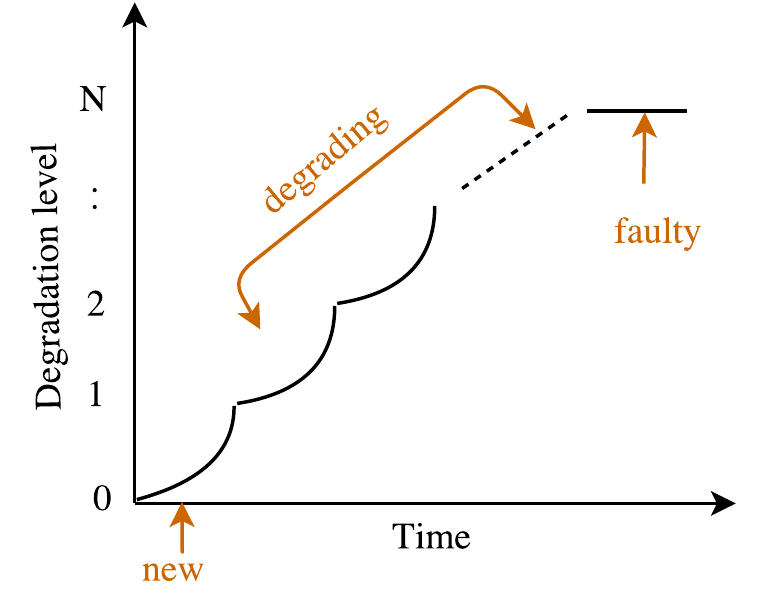}
			\caption{Timing diagram of degradation within an EBE. }
			\label{fig:EBETiming}
		\end{figure}
		\item \textit{Rate Dependency Events} - A new gate, introduced in ~\cite{ruijters2016fault} and labelled as \textit{RDEP}, accelerates the degradation rates of {$n$ dependent child nodes} and is depicted in Figure~\ref{fig:RDEPGate}.
		When the component connected to the input of the RDEP fails, the degradation rate of the dependent components is accelerated  with an acceleration factor $\gamma$. The corresponding timing diagram is shown in Figure~\ref{fig:RDEPTiming}. When the input signal is enabled ($input = 1$), the child EBE moves to the next degradation levels at a faster rate.  
		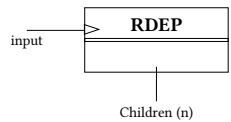
\begin{figure}[ht!]
			\centering
			\resizebox{3cm}{!}{ \begin{tikzpicture}[
      >=latex',
      auto
    ]
  \tikzstyle{int}=[draw,minimum size=2.5em,text centered,text width=3.5cm]
  \tikzstyle{intg}=[draw,minimum size=2.5em,text centered,text width=3.5cm]
     \node [int]  (ki1) [node distance=1.5cm] {\LARGE \textbf{RDEP}};
      \node [intg] (ki3) [node distance=.8cm,below of=ki1] { };
      \draw (-1.875,0) node[anchor=north]{}
  -- (-1.875,-.25) node[anchor=north]{}
  -- (-1.5,-0.125) node[anchor=south]{}
  -- cycle;
     \draw (-1.875,-0.15) node[anchor=north]{}
  -- (-3.375,-0.15) node[anchor=north]{\large input}
  -- cycle;
       \draw (0,-1.1) node[anchor=north]{}
  -- (0,-2) node[anchor=north]{\large Children (n)}
  -- cycle;
  
    \end{tikzpicture}}
			\caption{{RDEP gate with 1 input and dependent components also known as children.}}
			\label{fig:RDEPGate}
		\end{figure}
		\begin{figure}[h!]
			\centering
			\vspace{-0.5cm}
			\includegraphics[width=0.35\columnwidth]{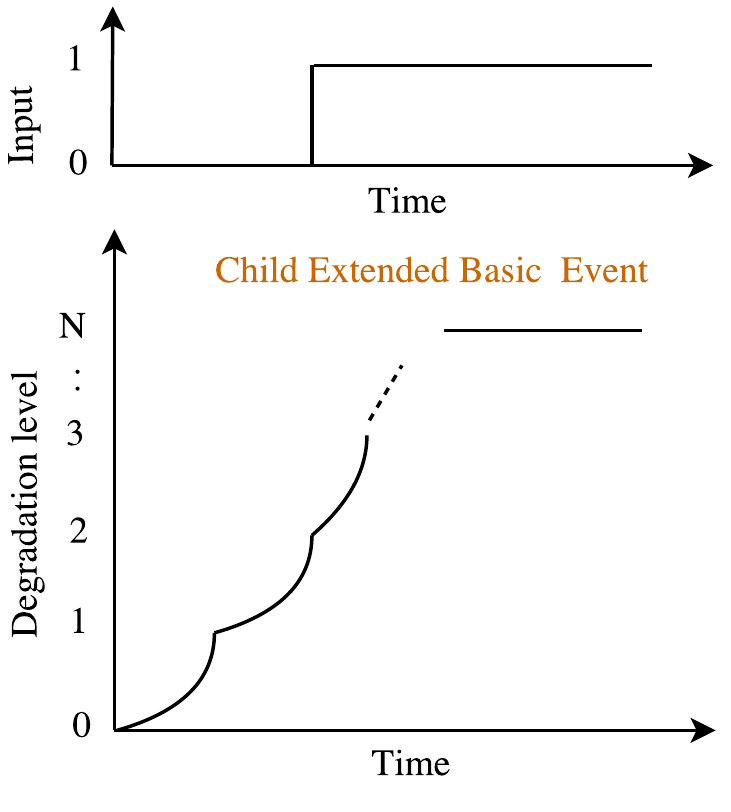}
			\caption{Degradation level evolution of child EBE showing effect of RDEP on degradation rate. {Note, when the input is equal to $1$ the curve representing the degradation rate to go from one degradation level to the next (e.g. going from degradation level 2 to 3) is steeper vs previous degradation level transitions (e.g. going from 0 to 1 or 1 to 2).} }
			\label{fig:RDEPTiming}
	`	\end{figure}

		\item \textit{Repair and Inspection modules} - 
		The repair module (RM) performs cleaning or replacements actions. These actions can be either carried out using fixed time schedules or when enabled by the inspection module (IM). The RM module performs periodic maintenance actions (clean or replace), independently of the IM. The IM performs periodic inspections and when components fall below a certain degradation threshold a maintenance action is initiated by the IM and performed by the RM (outside of the RM's periodic maintenance cycle).  The IM and RM modules are depicted in Figure~\ref{fig:IMRM}. The effect of performing a maintenance cleaning or replacement action on the EBE is illustrated in Figure~\ref{fig:RMIMtiming}. When a cleaning action is performed the EBE moves back to its previous degradation level, while when a replacement is performed the EBE moves back to the initial level.
		\begin{figure}[h!]
			\centering
			\includegraphics[width=0.3\columnwidth]{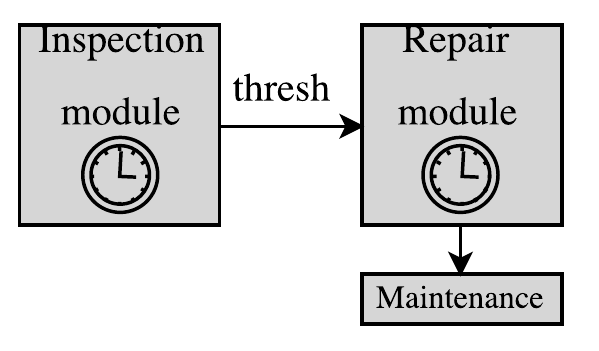}
			\caption{High-level description of the inspection and repair modules. {The repair module performs maintenance actions periodically (clean or replace). The inspection module performs inspections periodically and when the degradation level of an EBE reaches \textit{thresh} level, it triggers the repair module to perform a maintenance action immediately. }}
			\label{fig:IMRM}
		\end{figure}
		\begin{figure}[h!]
			\centering
			\includegraphics[width=0.8\columnwidth]{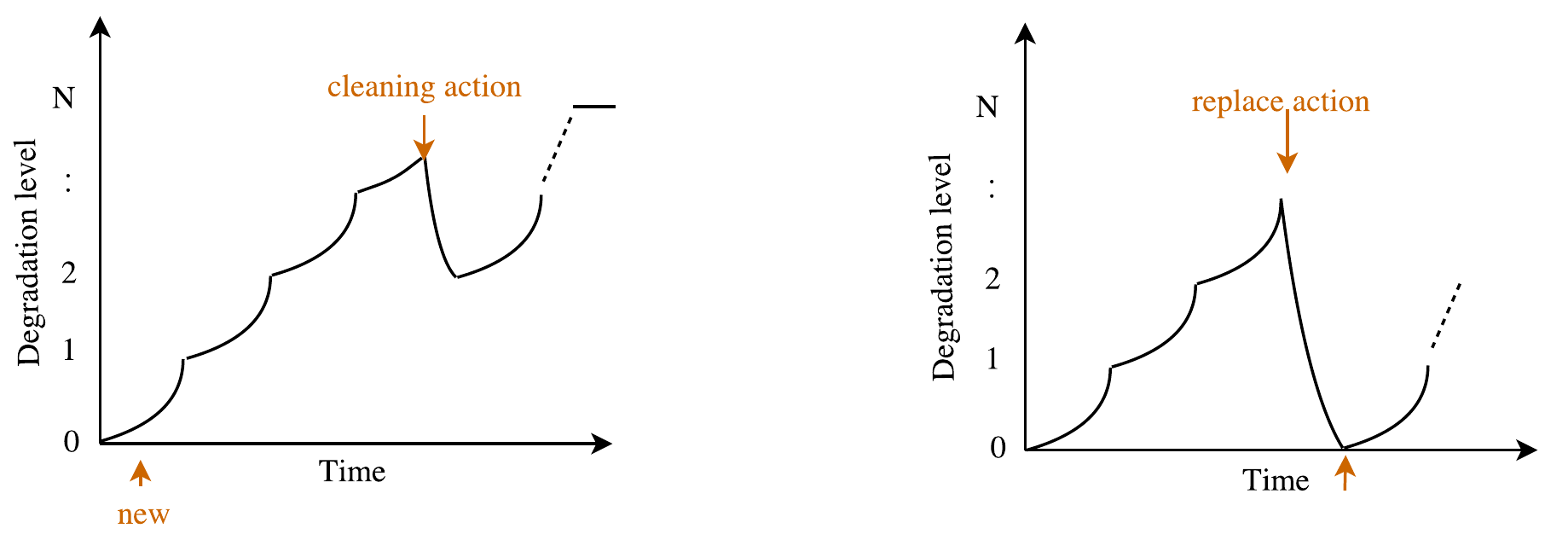}
			\caption{Degradation level progression of EBE for different maintenance actions.}
			\label{fig:RMIMtiming}
		\end{figure}
	\end{enumerate}
	A visual rendering of an FMT is given in Figure~\ref{fig:FMTsmall}. It is composed of five EBEs located at the bottom of the tree, one RDEP with one dependent child, three gates, one repair and inspection module and three events which show the different fault stages.

%
	\subsection{Probabilistic model checking}
	\label{subsec:Prelim:PMC}
	Model checking~\cite{Clarke86automaticverification} 
	is a well-established formal verification technique used to verify the correctness of finite-state systems. Given a formal model of the system to be verified in terms of labelled state transitions and the properties to be verified in terms of temporal logic, the model checking algorithm exhaustively and automatically explores all the possible states in a system to verify if the property is satisfiable or not. 
	\emph{Probabilistic model checking} (PMC) deals with systems that exhibit stochastic behaviour
	and is based on the construction and analysis of a probabilistic model of the system. We make use of CTMCs{, having both transition and state labels, } to perform stochastic modelling. Properties are expressed in the form of 
	Continuous Stochastic Logic (CSL)~\cite{kwiatkowska2007stochastic}, a stochastic variant of the well-known Computational Tree Logic (CTL)~\cite{Clarke86automaticverification} 
	which includes reward formulae. {Note, a system can be modelled using multiple CTMCs which represent different sub-components within the whole. Transition labels are then used to synchronise the individual CTMCs representing different parts of a system and in turn obtain the full CTMC representing the whole system.}
	
	 \begin{definition}[] \label{def:CTMC}
		{The tuple $C =(S, s_0, \Act, \Ap, L, \Rt)$ defines a CTMC which is composed of a set of states $S$, the initial state $s_0$, a finite set of transition labels $\Act$, a finite set of atomic propositions $\Ap$, a labelling function $L: S\rightarrow 2^{AP}$ and  the transition rate matrix $ \Rt: S \times  S \rightarrow \Rt_{\geq 0}$. }The rate $\Rt(s,s')$ defines the delay before which a transition between states $s$ and $s'$ takes place. If $\Rt(s,s') \ne 0$ then the probability that a transition between the states $s$ and $s'$ is 
		defined as $1 - e^{-{{\Rt}}(s,s') t} $ where $t$ is time. No transitions will  trigger if $\Rt(s,s') = 0$. 
		
	\end{definition}
	{The logic of  CSL specifies state-based properties for CTMCs, built out of propositional logic (with atoms $a \in \Ap$)
	, a steady-state operator ($\so$) that refers to the stationary probabilities, and a probabilistic operator ($\po$) for reasoning about transient state probabilities. The state formulas are interpreted over states of a CTMC, whereas the path formulas are interpreted over paths in a CTMC. 
	The syntax of CSL is:
	\begin{align*}
	\Phi &::= \texttt{true}\;|\;a\;|\;\Phi \wedge \Phi\;|\; \neg \Phi \;|\; \so_{\sim p}[\Phi] \;|\;\po_{\sim p}[\phi]\\
	\psi &::= \mathrm{X}\; \phi \;|\;\Phi\; \mathrm{U}^{\ge T} \;\Phi 
	\end{align*}
	where $\sim \in \{<,\le, =,\ge,>\}$, $p \in [0,1]$, $T \in \Rl_{\ge 0}$ is the time horizon, $\mathrm{X}$ is the next operator and $\mathrm{U}$ is the until operator. 
	The semantics of CSL formulas is given in~\cite{kwiatkowska2007stochastic}.} $S_{\sim p}[\Phi]$  asserts  that the steady-state probability for a $\Phi$-state meets the bound $\sim p$ whereas $P_{\sim p}[\Phi \;\mathrm{U}^{\le t} \;\Phi]$ asserts that with probability $\sim p$, by the time $t$ a state satisfying $\Phi$ will be reached such that all  preceding states satisfy $\Phi$.{ Additional properties can be specified by adding the notion of rewards. }The extended CSL logic adds reward operators, a subset of which are~\cite{kwiatkowska2007stochastic}:
	\begin{align*}
	\ro_{\sim r} [\co^{\le T}] \;|\; 
	 \ro_{\sim r} [\fo \;\Phi] 
	\end{align*}
	where $r,t \in \Rl_{\le 0}$ and $\Phi$ is a CSL formula. A state $s$ satisfies $\ro_{\sim r} [\co^{\le T}]$ 
	if, from state s, the expected reward cumulated up until $T$ time units have elapsed satisfies $\sim r$ and 
	$\ro_{\sim r} [\fo\; \Phi] $ is true if, from state $s$, the expected reward cumulated before a state satisfying $\Phi$ is reached  meets the bound $\sim r$.
	
	Examples of a CSL property with its natural language translation are: (i) $ \po_{\geq 0.95} [\fo ~complete]$ - ``The probability of the system eventually completing its execution successfully is at least 0.95". Each state (and/or transition) of the model is assigned a real-valued reward, allowing queries such as: 
	(ii) $\ro_{=?} [\fo ~success] $ - ``What is the expected reward accumulated before the system successfully terminates?" Rewards can be used to specify a wide range of measures of interest, for example, the total operational costs and the total percentage of time {during which} the system is available.
	
	
	\section{Formalizing FMTs using CTMCs}
	\label{subsec:FMTFrame:Modelling}
	
	\subsection{FMT Syntax}
	\label{subsubsec:FMTFrame:Modelling:Syntax}
	To formalise the syntax of FMTs using CTMCs, we first define the set \F, characterizing each FMT element by {type, inputs and  rates}. We introduce a new element called DELAY which will be used to model the deterministic time delays required by the extended basic events (EBE), repair module (RM) and inspection module (IM).  We restrict the set \F  to contain the  EBE, RDEP gate, OR gate, DELAY, RM and IM modules since these will be the components used in the case study presented in Section \ref{sec:CaseStudy}.
	
	\begin{definition}[]\label{def:FMTSem}
		The set \F $=\{EBE, RDEP, OR, DELAY,RM, IM\}$ of FMT elements consists of the following tuples. Here, $n, N \in \Na$ are natural numbers, $\mathit{thresh}, in, \mathit{trig} \in \{0,1\}$ take binary values, {$\Tcln$, $\Trpl,$ $\Trp,\Toh$, $\Tin \in \Rl_{\ge 0}$ are deterministic delays,  $\Td \in \Rl_{\ge 0}$ is a rate and $\gamma \in \Rl_{\ge 0}$ is a factor}.
		\begin{itemize}		
			\item $(EBE,\Td,\Tcln,\Trpl,N)$ represent the extended basic events  with $N$ discrete degradation levels, each of which degrade with a time delay equal to $T_{deg}$. It also takes as inputs the time taken to restore the EBE to the previous degradation level $\Tcln$ when cleaning is performed and the time taken to restore the EBE to its initial state $\Trpl$ following a replacement action.
			\item $(RDEP,n,\gamma,in, \Td)$ represents the RDEP gate with $n$ dependent children, acceleration factor $\gamma$, the input $in$ which activates the gate and $\Td$ the degradation rate of the dependent children.
			\item $(OR,n)$ represents the OR gate with $n$ inputs. {When either one of the inputs reaches the state labelled with $failed$, the OR gate returns a true signal.}
			\item $(RM,n,\Trp,\Toh,\Tin,\Tcln, \Trpl,{\mathit{thresh}}, \mathit{trig})$ represents the RM module which acts on $n$ EBEs (in our case, this corresponds to all the EBEs in the FMT).	The RM can either be triggered periodically to perform a cleaning action, every $\Trp$ delay, or a replacement action, every $\Toh$ delay, or by the IM when the delay $\Tin$ has elapsed and the $\mathit{thresh}$ condition is met. The time to perform a cleaning action is $\Tcln$, while the time taken to perform a replacement is $\Trpl$. The $\mathit{trig}$ signal ensures that when the component is not in the degraded states, no unnecessary maintenance actions are carried out. 
			\item $(IM,n,\Tin,\Tcln,\Trpl,\mathit{thresh})$ represents the IM module which acts on $n$ EBEs (in our case, this corresponds to all the EBEs in the FMT). The IM initiates a repair depending on the current state of the EBE. Inspections are performed in a periodic manner, every $\Tin$. If during an inspection, the current state of the EBE does not correspond to the \textit{new} or \textit{failed} state (i.e. the degradation level of the inspected EBE is below a certain threshold), the $thresh$ signal is activated and is sent to the RM. Once a cleaning action is performed the IM moves back to the initial state with a delay equal to  $\Tcln$ or $\Trpl$ depending on the maintenance action performed. 
			\item $(DELAY, T, N)$ represents the DELAY module which takes two inputs representing the deterministic delay $T \in \{\Td,$ $\Tcln,\Trpl,\Trp,\Toh,\Tin\}$ to be approximated using an Erlang distribution with $N$ states. 
			This DELAY module can be extended 
			by inclusion of a reset transition label, which when triggered restarts the approximation of the deterministic delay before it has elapsed. The extended DELAY module is referred to as  $(DELAY, T, N)_{ext}$.
		\end{itemize}
	\end{definition}
	The FMT is defined as a special type of directed acyclic graph $G=(V,E)$ where the vertices $V$ represent the gates and the events which represent an occurrence within the system, typically the failure of a subsystem down to an individual component level, and the edges $E$ which represent the connections between vertices. The vertices $V$ are labelled instances of elements in \F ~i.e. $V$ may contain multiple elements of the same component obtained from the set \F which are identified by their common element label. Events can either represent the EBEs or \textit{intermediate} events which are caused by one or more other events. The event at the top of the FMT is the top event (TE) and corresponds to the event being analysed -  modelling the failure of the (sub)system under consideration. The EBE are the leaves of the DAG. 
	For $G$ to be a well-formed FMT, we take the following assumptions 
	(i) vertices are composed of the OR, RDEP gates, (ii) there is only one top event, (iii) RDEP can only be triggered by EBEs and (iv) RM and IM are not part of the DAG tree but are modelled separately 
	This DAG formulation allows us to propose a framework in Subsection~\ref{subsec:FMTFrame:Dec} such that we can efficiently perform probabilistic model checking. 
	
	\begin{definition}[] \label{def:FMT}
		A fault maintenance tree is a directed acyclic graph 
		\noindent $G = (V,E)$ composed of vertices $V$ and edges $E$.	
	\end{definition}
	
	\subsection{Semantics of FMT elements }
	\label{subsubsec:FMTFrame:Modelling:SemanticsElements}
	
	Next, we provide the semantics for each FMT element, which are composed using the syntax of CTMC (cf. Definition~\ref{def:CTMC}). 
	These elements are then instantiated based on the underlying FMT structure to form the semantics of the whole FMT.{ We obtain the semantics of the whole FMT via synchronisation of transition labels between the different CTMCs representing the individual FMT elements. This is further explained in subsection~\ref{subsubsec:FMTFrame:Modelling:Semantics}}. 
	
	\paragraph{\textbf{DELAY}}
	\label{par:FMTFrame:Modelling:Clck}
	{We define the semantics for the $(DELAY,T,N)$ element 
		using Figure~\ref{fig:ClkNoRst} and describe the  corresponding CTMC  
		using the set of states given by  $D =\{d_0,d_1,\dots, d_{N+1}\}$, the initial state $d_0$, the set of transitions labels  $\Act=\{\texttt{trigger}, \texttt{move}\}$, the set of atomic propositions $\Ap =\{T\}$ with 
		$L(d_0)  = \dots =L(d_{N}) = \emptyset$, and $L(d_{N+1}) = \{T\}$. The rate matrix $\Rt$ becomes clear from Figure~\ref{fig:ClkNoRst} and  
		\begin{align}
		\Rt_{ij} = \begin{cases}
		\mu &  i= 0 \wedge j =1,\\
		\frac{N}{T} & ((i \ge 1 \vee i < N+1)  \wedge  j = i+1) \vee (i=N+1 \wedge j=1), \\
		0 & \text{otherwise},
		\end{cases}
		\end{align}
		with $i$ representing the current state, $j$ is the next state and $\mu$ is a fixed large value corresponding to introducing a negligible delay, which is used to trigger all the DELAY modules at the same time (cf. Definition~\ref{def:CTMC}).  }
	In Figure~\ref{fig:ClkRst} we define the semantics of $(DELAY,T,N)_{ext}$. This results in the CTMC described using the state space $D =\{d_0,d_1,\dots, d_{N+1}\}$, the initial state $d_0$, the set of transition labels $\Act=\{\texttt{trigger},$ $ \texttt{move}, \texttt{reset}\}$, the set of 
	atomic propositions $\Ap =\{T\}$, the labelling function $L(d_0) = L(d_1) = \dots =L(d_{N}) = \emptyset$, and $L(d_{N+1}) = \{T\}$ and the rate matrix $\Rt$ where \begin{align}
	\Rt_{ij} = \begin{cases}
	\mu &  i= 0 \wedge j =1,\\
	1           & (i \ge 2 \vee i < N+1) \wedge  j = 1,\\
	\frac{N}{T} & ((i \ge 1 \vee i < N+1)  \wedge j = i+1) \vee (i=N+1 \wedge j=1), \\
	0 & \text{otherwise},
	\end{cases}
	\end{align}
	with $i$ representing the current state and $j$ is the next state. In both instances, the deterministic delays is approximated using an Erlang distribution
	~\cite{hoque2015towards} and all DELAY modules are synchronised to start together using the \texttt{trigger} transition label.  The extended DELAY module have the transition labels \texttt{reset} which restarts the Erlang distribution approximation whenever the guard condition is met at a rate of $1 \times \Rt_{sync}$ where $\Rt_{sync}$ is the rate coming from the use of  synchronisation with other modules causing the reset to occur ( as explained in Subsection~\ref{subsubsec:FMTFrame:Modelling:Semantics}). This is required when a maintenance action is performed which restores the EBE's state back to the original state and thus restart the degradation process, before the degradation time has elapsed. 
	\begin{remark} \label{remark:Er}
		A random variable $Z \in \Rl_{+}$ has an Erlang distribution with 	$k \in \Na$ stages and a rate $\lambda \in \Rl_{+}, Z \sim Erlang(k,\lambda)$, if $Z = Y_1 + Y_2 + \dots Y_k$ where each $Y_i$ is exponentially distributed with rate $\lambda$. The cumulative density function of the Erlang distribution is characterised using
		\begin{equation} \label{eqn:cdf}
		f(t;k,\lambda) = 1- \sum_{n=0}^{k-1}\frac{1}{n!} \exp(-\lambda t)(\lambda t)^n \quad \textrm{for } t,\lambda \ge 0,
		\end{equation}
		and for $k=1$, the Erlang distribution simplifies to the exponential distribution. In particular, the sequence $Z_k \sim Erlang(k,\lambda k)$ converges to the deterministic value $\frac{1}{\lambda}$ for large $k$. Thus, we can approximate a deterministic delay $T$ with a random variable $Z_k \sim Erlang(k,\frac{k}{T})$~\cite{bortolussi2012fluid,hoque2014probabilistic}. Note, there is a trade-off between the accuracy and the resulting blow-up in size of the CTMC model for larger values of $k$ (a factor of $~k$ increase in the model size)~\cite{hoque2015towards}. In this work, the Erlang distribution will be used to model the fixed degradation rates, the maintenance and inspection signals. This is a similar approach taken in ~\cite{ruijters2016fault} where degradation phases are approximated by an (k,$\lambda$)-Erlang distribution.
	\end{remark}
	\begin{figure}[ht!]
		\centering
		\psfragscanon
			\resizebox{.75\textwidth}{!}{\begin{tikzpicture}[font=\sffamily]

\tikzset{node style/.style={state, 
		minimum width=1.2cm,
		line width=0.5mm,
		fill=white!20!white}}

\node[node style, initial above] at (0, 0)  (s0)     {\Large $d_0$};
\node[node style] at (3, 0)  (s1) {\Large $d_1$};
\node[node style] at (6, 0)  (s2) {\Large $d_2$};
\node[node style] at (9, 0)  (s3) {\Large $d_3$};
\node[]           at (12, 0) (s4) {\Large $\dots$};
\node[node style] at (15, 0) (s5) {\Large $d_{N+1}$};
\draw[every loop,
auto=right,
line width=0.5mm,
>=latex,
draw=black,
fill=black]
(s0)     edge[ auto=left] node {\texttt{trigger},$\mu$} (s1)
(s1)     edge[bend left=20, auto=left] node {\Large\texttt{move},$\frac{N}{T}$} (s2)
(s2)     edge[bend left=20, auto=left] node {\Large\texttt{move},$\frac{N}{T}$} (s3)
(s3)     edge[bend left=20, auto=left] node {\Large\texttt{move},$\frac{N}{T}$} (s4)
(s4)     edge[bend left=20, auto=left] node {\Large\texttt{move},$ \frac{N}{T}$} (s5)
(s5)     edge[bend left=32, auto=right] node {\Large{\texttt{move}},\Large$\frac{N}{T}$} (s1);
\end{tikzpicture}}
			\caption{{CTMC representing DELAY with $N$ states used to approximate a delay equal to $T$ approximated using $Erlang(N,\frac{N}{T})$. The transition labels $\Act =\{ \texttt{trigger}, $ $\texttt{move}\}$ are shown on each of the transitions. The state labels are not shown and the initial state of the CTMC is pointed to using an arrow labelled with start. }}\label{fig:ClkNoRst}
			\resizebox{.75\textwidth}{!}{\begin{tikzpicture}[font=\sffamily]

\tikzset{node style/.style={state, 
		minimum width=1.2cm,
		line width=0.5mm,
		fill=white!20!white}}

\node[node style, initial above] at (0, 0)  (s0)     {\Large $d_0$};
\node[node style] at (3, 0)  (s1)     {\Large $d_1$};
\node[node style] at (6, 0)  (s2) {\Large $d_2$};
\node[node style] at (9, 0)  (s3) {\Large $d_3$};
\node[]           at (12, 0) (s4) {\Large $\dots$};
\node[node style] at (15, 0) (s5) {\Large $d_{N+1}$};

\draw[every loop,
auto=right,
line width=0.5mm,
>=latex,
draw=black,
fill=black]
(s0)     edge[ auto=left] node {\texttt{trigger} ,$\mu$} (s1)
(s1)     edge[bend left=20, auto=left] node {\Large\texttt{move} ,$\frac{N}{T}$} (s2)
(s2)     edge[bend left=20, auto=left] node {\Large\texttt{move} ,$\frac{N}{T}$} (s3)
(s3)     edge[bend left=20, auto=left] node {\Large\texttt{move} ,$\frac{N}{T}$} (s4)
(s4)     edge[bend left=20, auto=left] node {\Large\texttt{move} ,$\frac{N}{T}$} (s5)
(s2)     edge[bend left=44, auto=right] node {\Large\texttt{reset},1} (s1)
(s3)     edge[bend left=46, auto=right] node {\Large\texttt{reset},1} (s1)
(s4)     edge[bend left=48, auto=right] node {\Large\texttt{reset},1} (s1)
(s5)     edge[bend left=48, auto=right] node {\Large\texttt{reset},1} (s1);
\end{tikzpicture}}
			\caption{{CTMC representing the extended DELAY with $N$ states used to approximate a delay equal to $T$. Delay approximated using $Erlang(N,\frac{N}{T})$. The transition labels $\Act =\{ \texttt{trigger}, $ $\texttt{move}, \texttt{reset}\}$ are shown on each of the state transitions, while the state labels are not shown. 
			}}\label{fig:ClkRst}
		\label{fig:Clk}
	\end{figure}

\begin{figure}[ht!]
		\centering
		\resizebox{.9\textwidth}{!}{\begin{tikzpicture}[font=\sffamily]

\tikzset{node style/.style={state, 
		minimum width=1.2cm,
		line width=0.5mm,
		fill=white!20!white}}

\node[node style, initial above] at (0, 0)  (s0)     {\Large $s_0$};
\node[node style] at (5, 0)  (s1)     {\Large $s_1$};
\node[node style] at (10, 0) (s2) {\Large $s_2$};
\node[node style] at (15, 0) (s3) {\Large $s_3$};
\draw[every loop,
auto=right,
line width=0.5mm,
>=latex,
draw=black,
fill=black]
(s0)     edge[bend left=20, auto=left] node {$\texttt{degrade}_1,\lambda$} (s1)
(s1)     edge[bend left=20, auto=left] node {$\texttt{degrade}_2,\lambda$} (s2)
(s2)     edge[bend left=20, auto=left] node {$\texttt{degrade}_3,\lambda$} (s3)
(s3)     edge[bend left=10, auto=right] node {$\texttt{perform\_clean},1$} (s2)
(s2)     edge[bend left=10, auto=right] node {$\texttt{perform\_clean},1$} (s1)
(s1)     edge[auto=right] node {$\texttt{perform\_clean},1$} (s0)
(s3)     edge[bend left=23, auto=right] node {$\texttt{perform\_replace},1$} (s0)
(s2)     edge[bend left=23, auto=right] node {$\texttt{perform\_replace},1$} (s0)
(s1)     edge[bend left=23, auto=right] node {$\texttt{perform\_replace},1$} (s0);
\end{tikzpicture}}
		\caption{{CTMC representing the EBE with $N =3$ with the transition labels $\Act_{EBE} =\{ \texttt{degrade}_{i \in \{1,2,3\}}, $ $\texttt{perform\_clean}, \texttt{perform\_replace}\}$  on each of the state transitions. For clarity, the state labels are not shown. 
		The deterministic delays contained represent the transition label that is triggered when the delay generated by the corresponding DELAY module has elapsed. The degradation rate is equal to $\lambda = \frac{N}{MTTF}$ where MTTF is the components mean time to failure.   }}
		\label{fig:EBE}
	\end{figure}
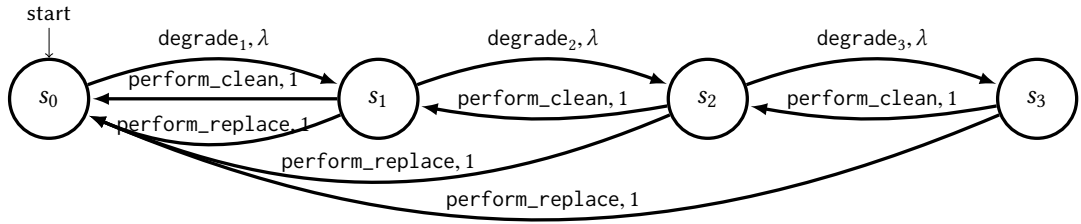

	\paragraph{\textbf{Extended Basic Events (EBE)}}
	\label{par:FMTFrame:Modelling:EBE}
	The EBE 
	are the leaves of the FMT and incorporate the component's degradation model. EBE are a function of the total number of degradation steps $N$ considered. 
	Figure \ref{fig:EBE} shows the semantics of the $(EBE,T_{deg}, T_{cln},T_{rep},N=3)$. The corresponding CTMC {is described by the tuple $(\{s_0,s_1,s_2,s_3\},$ $ s_0,\Act_{EBE},$  $\Ap_{EBE}, L_{EBE}, \Rt_{EBE})$ where  $s_0$ is the initial state , \begin{align*}
		\Act_{EBE} &=\{ \texttt{degrade}_{i\in\{0,\dots,N\}}, \texttt{perform\_clean},\texttt{perform\_replace}\},
		\end{align*} the atomic propositions $\Ap_{EBE} = \{\mathit{new},\mathit{thresh},$ $ \mathit{failed}\}$, the labelling function $L(s_0) = \{new\},$ $ L(s_1) = L(s_2) = \{thresh\}, L(s_3) = \{failed\}$ }and 
		\begin{align*}
	\Rt_{EBE}= \left[\begin{matrix}
	0&1&0&0\\
	1&0&1&0\\	
	1&1&0&1\\
	1&0&1&0\\
	\end{matrix}\right].
	\end{align*}

	{The deterministic time delays taken as inputs are modelled using three different DELAY modules:
		\begin{enumerate}
			\item an extended DELAY module approximating $\Td$ with the transition label \texttt{move} replaced with \texttt{degrade}$_N $such that synchronisation between the two CTMCs is performed (explained in Subsection~\ref{subsubsec:FMTFrame:Modelling:Semantics}).  When $\Td$ has elapsed the transition labelled with 
			\texttt{degrade}$_N$ is triggered and the EBE moves to the next state at a rate \footnote{This is a direct consequence of synchronisation and corresponds to $\Rt \times \Rt_{EBE}$. Refer to Subsection~\ref{subsubsec:FMTFrame:Modelling:Semantics}}  equal to $\frac{N}{\Td} \times 1$. The \texttt{reset} transition label and the corresponding transitions are replicated in extended DELAY module and replaced with \texttt{perform\_clean} and \texttt{perform\_replace}. When the the previous state (if cleaning action is carried out) or to the initial state (if replace action is performed).
			\item a DELAY module approximating $\Tcln$  with the transition label \texttt{move} replaced with \texttt{perform\_clean}.  When $\Tcln$ has elapsed the transition with transition label \texttt{perform\_clean} is triggered and the EBE moves to the previous state at a rate  equal to $\frac{N}{\Tcln}$. 
			\item a DELAY module approximating $\Trpl$  with the transition label \texttt{move} replaced with \texttt{perform\_replace}.  When $\Trpl$ has elapsed the transition label \texttt{perform\_replace} is triggered and the EBE moves to the initial state at a rate  equal to $\frac{N}{\Trpl}$. 
		\end{enumerate}
		The transition labels \texttt{perform\_clean} and \texttt{perform\_replace} cannot be triggered at the same time and it is assumed that $\Tcln \neq \Trpl$. This is a realistic assumption as only one maintenance action is performed at the same time. } 	
	
	\paragraph{\textbf{RDEP gate}}
	\label{par:FMTFrame:Modelling:RDEP}
	The RDEP gate has static semantics and is used in combination with the semantics of its $n$ dependent EBEs. When triggered ($input=1$), the associated EBE reaches the state labelled $\mathit{failed}$, the degradation rate of the $n$ dependent children is accelerated by a factor $\gamma$. We model the $input$ signal using
	\begin{align}\label{eqn:in}
	input &= \begin{cases}
	1 & L(s) = \mathit{failed},\\
	0 & \text{otherwise},\\
	\end{cases}
	\end{align}
	where $L(s)$ is the label of the  current state of the associated EBE (cf. Figure~\ref{fig:RDEPTiming}). 
	{ Similarly, we map the RDEP gate function using
		\begin{align}\label{eqn:rdep}
		RA &= \begin{cases}
		\gamma T_{deg_1},\dots, \gamma T_{deg_n}& input= 1,\\
		T_{deg_1},\dots, T_{deg_n} & \text{otherwise},\\
		\end{cases}
		\end{align}
		where $T_{deg_i}, i \in {1, \dots n}$ corresponds to the degradation rate of the $n$ dependent children.}   \footnote{Note, this effectively results in changing the deterministic delay being modelled by the DELAY module to a new value if $input = 1$.} %

\begin{figure}[ht!]
		\psfragscanon
		\centering
		\resizebox{.45\textwidth}{!}{    \begin{tikzpicture}[font=\sffamily]
 
        \tikzset{node style/.style={state, 
                                    minimum width=1.2cm,
                                    line width=0.5mm,
                                    fill=white!20!white}}
 
        \node[node style, initial left] at (0, 0)  (s0)     {\Large $rm_0$};
        \node[node style] at (6, 0)  (s1)     {\Large $rm_1$};

        \draw[every loop,
              auto=right,
              line width=0.5mm,
              >=latex,
              draw=black,
              fill=black]
            (s0)     edge[loop above, auto=left] node {\texttt{inspect},thresh =0,$1$ } (s0)
             (s0)     edge[loop below, auto=left] node {\texttt{check\_maintenance}, trig =0,$1$ } (s0)
            (s0)     edge[bend left=20, auto=left] node {\texttt{check\_maintenance}, trig=1,$1$} (s1)
            (s0)     edge[bend left=00, auto=left] node {\texttt{inspect}, thresh =1,$1$} (s1)
            (s1)     edge[bend left=20, auto=left] node {\texttt{trigger\_maintenance},$1$} (s0);
           
    \end{tikzpicture}}
		\caption{{CTMC representing the RM with $\Act_{RM} = \{ \texttt{inspect}, \texttt{check\_maintenance}, \texttt{perform\_maintenance} \}$ shown on the state transitions. The guard condition $\mathit{trig} = 0 / 1$ or $\mathit{thresh} = 0/1$ must be satisfied for the corresponding transition to trigger when it is activated via synchronisation with the transition label.}}\label{fig:RM}
	\end{figure}
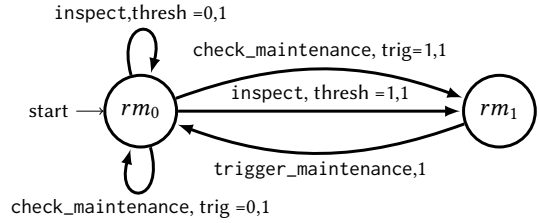

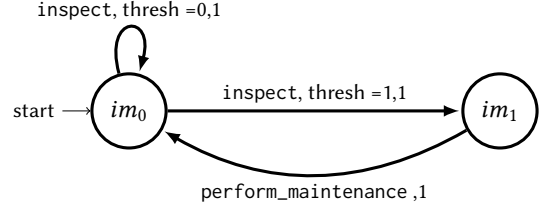
\begin{figure}[ht!]
		\psfragscanon
		\centering		\resizebox{.45\textwidth}{!}{    \begin{tikzpicture}[font=\sffamily]
 
        \tikzset{node style/.style={state, 
                                    minimum width=1.2cm,
                                    line width=0.5mm,
                                    fill=white!20!white}}
 
        \node[node style, initial left] at (0, 0)  (s0)     {\Large $im_0$};
        \node[node style] at (6, 0)  (s1)     {\Large $im_1$};

        \draw[every loop,
              auto=right,
              line width=0.5mm,
              >=latex,
              draw=black,
              fill=black]
            (s0)     edge[loop above, auto=left] node {\texttt{inspect}, thresh =0,$1$} (s0)
            (s0)     edge[bend left=00, auto=left] node {\texttt{inspect}, thresh =1,$1$} (s1)
            (s1)     edge[bend left=30, auto=left] node {\texttt{perform\_maintenance} ,$1$} (s0);
           
    \end{tikzpicture}}
			\caption{{CTMC representing the IM with $\Act_{IM} = \{\texttt{inspect}, \texttt{perform\_maintenance}\}$ shown on the state transitions. The guard condition $\mathit{trig} = 0 $ and $\mathit{thresh} = 1$ must be satisfied for the corresponding transition to trigger when it is activated via synchronisation with the transition label.}}\label{fig:IM}
	\end{figure}

	\paragraph{\textbf{OR gate}}
	\label{par:FMTFrame:Modelling:OR}
	The OR gate 
	indicates a failure when either of its input nodes have failed and also does not have semantics itself but is used in combination with the semantics of its $n$ dependent input events  (EBEs or intermediate events). 
	We use
	{\begin{align}\label{eqn:OR}
	FAIL &= \begin{cases}
	0    & E_{1} = 1 \wedge \dots \wedge E_{n} = 1, \\
	1    & \text{otherwise},\\
	\end{cases}
	\end{align}}
	where $E_{i} = 1, i \in 1 \dots n$ corresponds to when the $n$ events (cf. Def.~\ref{def:event}), connected to the OR gate, represent a failure in the system. In the case of EBEs, $E_1 =1$ occurs when the EBE reaches the $\mathit{failed}$ state .
	\paragraph{\textbf{Repair module (RM)}}
	\label{par:FMTFrame:Modelling:RM}
	Figure  ~ \ref{fig:RM} shows the semantics of $(RM,n,$ $\Trp, \Toh, \Tin,$ $ \Tcln,$ $\Trpl,$ $\mathit{thresh},\mathit{trig})$.    The CTMC is described using { the state space  $\{rm_0,rm_1\}$, the initial state $rm_0$, the transition labels
	\begin{align*}
	 \Act_{RM} = \{\texttt{inspect}, \texttt{check\_clean}, \texttt{check\_replace},\texttt{trigger\_clean}, \texttt{trigger\_replace} \},
	\end{align*}
 the atomic propositions $\Ap=\{maintenance\}$, the labelling function $L(rm_{0})=\{ \emptyset \},$ $ L(rm_1) = \{ \mathit{maintenance} \}$ and with }	
\begin{align*}
	\Rt_{IM} = \left[\begin{matrix}
	1&1\\
	1&0\\	
	\end{matrix}\right].
	\end{align*} 
	{For brevity in Figure~\ref{fig:RM}, we used the transition labels \texttt{check\_maintenance} and \texttt{trigger\_maintenance}. The transition label \texttt{check\_maintenance}  and corresponding transitions are replicated and the transition labels replaced by  \texttt{check\_clean}  or \texttt{check\_replace} to allow for both type of maintenance checks.  Similarly, the transition label \texttt{trigger\_maintenance} and corresponding transitions are duplicated and the transition labels replaced by  \texttt{trigger\_clean}  or \texttt{trigger\_replace} to allow the initiation of both type of maintenance actions to be performed. Due to synchronisation, only one of the transitions may trigger at any time instance (as explained in Subsection~\ref{subsubsec:FMTFrame:Modelling:Semantics}).	
		\noindent The transition labels   \texttt{trigger\_clean}  or \texttt{trigger\_replace} correspond to the transition label \texttt{trigger} within the DELAY module approximating the deterministic delays $\Tcln$ and $\Trpl$ respectively. 
		 The deterministic delays which trigger \texttt{inspect}, \texttt{check\_clean}  or \texttt{check\_replace}  correspond to when the time delays $\Tin, \Trp$ and $\Toh$ respectively, have elapsed. All these signals are generated using individual DELAY modules with the \texttt{move} transition label for each module replaced using   \texttt{inspect}, \texttt{check\_clean}  or \texttt{check\_replace} respectively.
}
	The $\mathit{thresh}$ signal is modelled using
	\begin{align}\label{eqn:thresh}
	\mathit{thresh} &= \begin{cases}
	1 &  L(s_{j,1}) = \mathit{thresh} \vee \dots \vee L(s_{j,n}) = \mathit{thresh},\\
	0   & \text{otherwise},\\
	\end{cases}
	\end{align}
	where $L(s_{j,i}), j \in 0 \dots N, i \in 1 \dots n$ correspond to  the label of the current state $j$ of each of the $n$ EBE. Similarly, we model the $\mathit{trig}$ signal using
	\begin{align}\label{eqn:trig}
	\mathit{trig} &= \begin{cases}
	1 &  L(s_{j,1}) \neq  \mathit{new} \vee \dots \vee L(s_{j,n}) \neq \mathit{new} ,\\
	0   & \text{otherwise}.\\
	\end{cases}
	\end{align}
	Both signals act as guards which when triggered determine which transition to perform (cf. Fig.~ ~ \ref{fig:RM}).
	\paragraph{\textbf{Inspection module (IM) }}
	\label{par:FMTFrame:Modelling:IM}
	The semantics of the $(IM,n,\Tin,$ $ \Tcln,\Trpl,$ 
	$\mathit{thresh})$ is depicted in Figure~~ \ref{fig:IM}.{ The CTMC is defined using the tuple $(\{im_0,im_1\},im_0,\Act_{IM}, \Ap_{IM}, L_{IM}, \Rt_{IM})$. Here, 
	\begin{align*}
	\Act_{IM} = \{\texttt{inspect}, \texttt{perform\_clean},\texttt{perform\_replace} \},
	\end{align*} $
	\Ap_{IM}= \{\emptyset\},$ with $	L(s_0) =L(s_1) = \emptyset$
	and 	
	\begin{align*}
		\Rt_{IM} = \left[\begin{matrix}
		1&1\\
		1&0\\	
		\end{matrix}\right].\end{align*}}The $\mathit{thresh}$ signal corresponds to same signal used by the RM, given using~\eqref{eqn:thresh}. In  Figure~\ref{fig:IM}, for clarity, we use the transition label \texttt{perform\_maintenance}. This transition label and corresponding transitions are duplicated and the transition labels are replaced by  either \texttt{perform\_clean}  or \texttt{perform\_replace} to allow for both type of maintenance actions to be performed when one of them is triggered using synchronisation.  The same DELAY modules used in the RM and EBE to represent the deterministic delays are used by the IM. The DELAY module used to represent the  deterministic delays $\Tcln$ and $\Trpl$ triggers the transition labels \texttt{perform\_clean}  or \texttt{perform\_replace}. This  represents that the maintenance action has completed.
	\subsection{{ Semantics of composed FMT}}
	\label{subsubsec:FMTFrame:Modelling:Semantics}	 
	{Next, we show how to obtain the semantics of a FMT from the semantics of its elements using the FMT syntax introduced in Subsection~\ref{subsubsec:FMTFrame:Modelling:Syntax}. We define the DAG $G$ by defining the vertices $V$ and  the corresponding events $E$. The leaves of the DAG are the events corresponding to the EBE. The events $E$ are connected to the vertices $V$, which trigger the corresponding auxiliary function used to represent the semantics of the gates. The $Events$ connected to the RM and IM are initiated by triggering the auxiliary functions  $\mathit{thresh}$ and $\mathit{trig}$ given using ~\eqref{eqn:thresh} and ~\eqref{eqn:trig} respectively. Based on the structure of $G$, we compute the corresponding CTMC by applying parallel composition of the individual CTMCs representing the elements of the FMT. The parallel composition formulae are derived from~\cite{hermanns2011concurrency} and defined as follows,}
	\begin{definition}[Interleaving Synchronization] \label{def:IntSynch}
		The interleaving synchronous product of $C_1 = (S_1,s_{01},$ $ \Act_{1},$ $  \Ap_1, L_1,\Rt_1)$ and $C_2 =(S_2,s_{02}, \Act_{2}, \Ap_2, L_2,\Rt_2)$ is $C_1 ||C_2 = (S_1 \times  S_2, (s_{01} , s_{02}),\Act_1 \cup \Act_2, \Ap_1 \cup \Ap_2, L_1 \cup L_2,\Rt)$ where $\Rt$ is given by:
		{\begin{align*}
			\frac{s_1 \xrightarrow{\alpha_1, \lambda_1} s_1'}{(s_1,s_2) \xrightarrow{\alpha_1,\lambda_1} (s_1',s_2)},~\text{and}~
			\frac{s_2 \xrightarrow{\alpha_2, \lambda_2} s_2'}{(s_1,s_2) \xrightarrow{\alpha_2,\lambda_2} (s_1,s_2')},
			\end{align*}
		}
	and $s_1,s_1' \in S_1$, $\alpha_1 \in \Act_1$, $\Rt_1(s_1,s_1') = \lambda_1$, $s_2,s_2' \in S_2$, $\alpha_2 \in \Act_2$, $\Rt_2(s_2,s_2') = \lambda_2$. 
	\end{definition}
	\begin{definition}[Full Synchronization]\label{def:FullSynch}
		The full synchronous product of $C_1 = (S_1,s_{01}, \Act_{1}, \Ap_1, L_1,\Rt_1)$ and $C_2 =(S_2,s_{02}, \Act_{2}, \Ap_2, L_2,\Rt_2)$ is $C_1 ||C_2 = (S_1 \times  S_2, (s_{01} , s_{02}),\Act_1 \cup \Act_2, \Ap_1 \cup \Ap_2, L_1 \cup L_2,\Rt)$ where $\Rt$ is given by:
		{\begin{align*}
		\frac{s_1 \xrightarrow{\alpha, \lambda_1} s_1' ~ \text{and}~ s_2 \xrightarrow{\alpha, \lambda_2} s_2' }{(s_1,s_2) \xrightarrow{\alpha,\lambda_1\times \lambda_2} (s_1',s_2')},
		\end{align*}
	}
	and $s_1,s_1' \in S_1$, $\alpha \in \Act_1 \wedge \Act_2$, $\Rt_1(s_1,s_1') = \lambda_1$, $s_2,s_2' \in S_2$, $\alpha_2 \in \Act_2$, $\Rt_2(s_2,s_2') = \lambda_2$. 
	\end{definition}
	
\begin{figure}[h!]
				\centering
				\includegraphics[width=.5\textwidth]{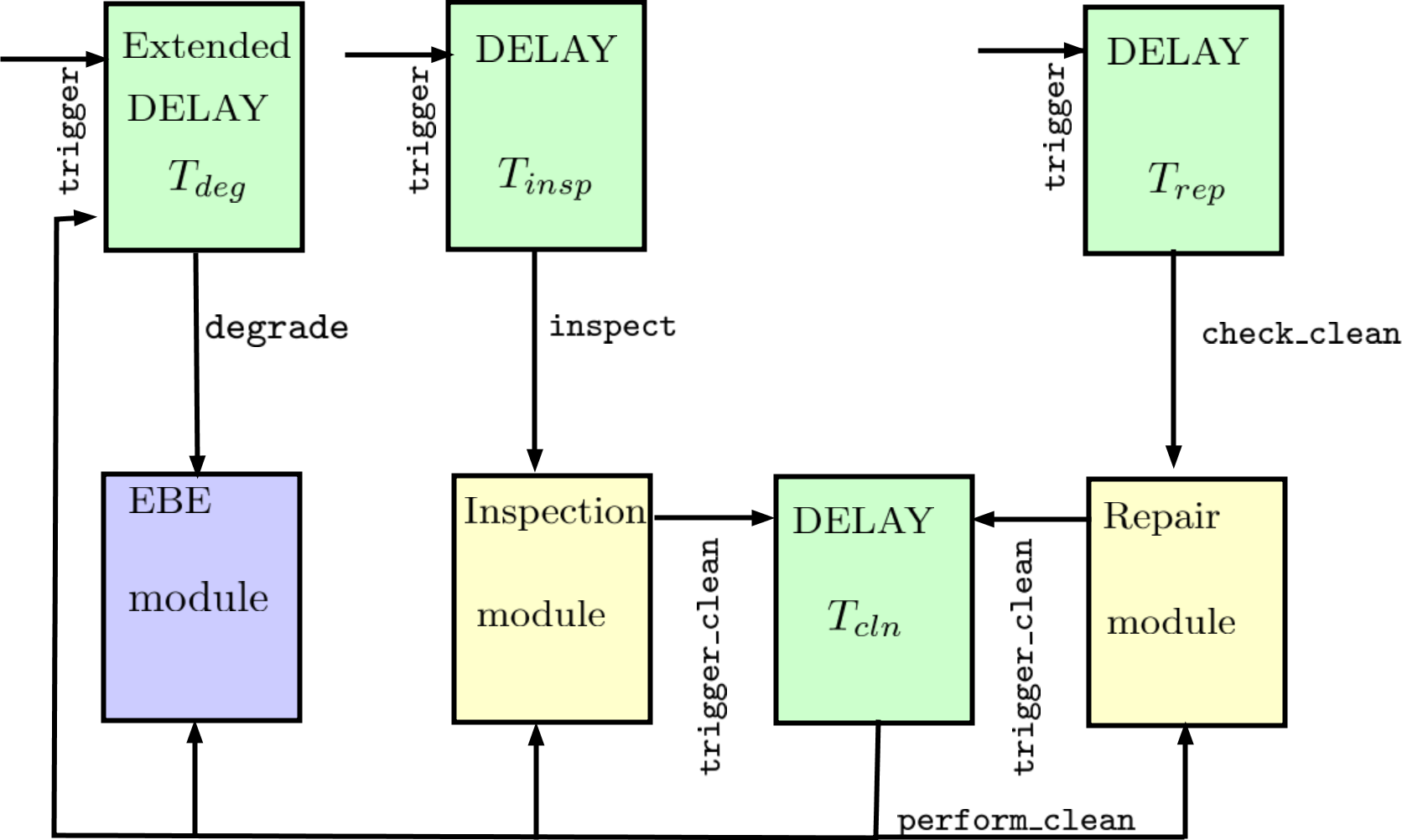}
				\caption{{Block diagram showing the synchronisation connections between one component and the other, together with the corresponding transition label which triggers synchronisation.}}
				\label{fig:Synch}
		\end{figure}

	For any pair of states, synchronisation is performed either using interleaving or full synchronisation. For full synchronisation, as in Definition~\ref{def:IntSynch}, the rate of a synchronous transition is defined as the product of the rates for each transition. The intended rate is specified in one transition and the rate of other transition(s) is {specified as one}. {For instance, the RM synchronises using full synchronisation with the DELAY modules representing $\Tin$, $\Trp$ and $\Trpl$ and therefore, to perform synchronisation between the RM and the DELAY modules, the rates of all the transitions of RM should have a value of one (cf. Fig.~~ \ref{fig:RM}), while the rate of the DELAY modules represent the actual rates (cf. Fig~\ref{fig:ClkNoRst} and Fig~\ref{fig:ClkRst}). The same principle holds for the EBEs and the IM. We refer the reader to Table~\ref{tab:synch} to further understand the synchronisation between the FMT components and the method employed for parallel composition.} 
	\begin{table*}[ht!]
		\centering
		\resizebox{\textwidth}{!}{
		\begin{tabular}{llll}
			\textbf{Component}& \textbf{Synchronised with component} & \textbf{Transition label}& \textbf{Synchronisation method}\\ \hline \hline
			DELAY representing $\Td$ &  DELAY modules representing $\Tcln,\Trpl, \Tin$ & \texttt{trigger} & Full synchronisation\\ 
			RM                       &  DELAY module representing $\Trp$ & \texttt{trigger\_clean} & Full synchronisation\\ 
			RM                       &  DELAY module representing $\Toh$ & \texttt{trigger\_replace} & Full synchronisation\\ 
			EBE & DELAY representing $\Td$ & \texttt{degrade}$_N$& Full synchronisation \\
			DELAY representing $\Tcln$  &  RM, EBE & \text{check\_clean} & Full synchronisation\\
			DELAY representing $\Trpl$  &  RM, EBE & \text{check\_replace} & Full synchronisation\\
			DELAY representing $\Tin$  &  RM, IM & \text{inspect} & Full synchronisation\\
			DELAY representing $\Trp$  &  RM, IM, EBE & \text{perform\_clean} & Full synchronisation\\
			DELAY representing $\Toh$  &  RM, IM, EBE & \text{perform\_replace} &Full synchronisation \\
			EBE                        & RM,IM, all DELAY modules, other EBEs & 	- & Interleaving synchronisation\\		\hline \hline
			 
		\end{tabular}}
	\caption{{Performing synchronisation between the different FMT components and the synchronisation method used.}}
	\label{tab:synch}
	\end{table*}
%
		Consider a simple example showing the time signals and synchronisations required for modelling an 
		EBE and the RM and IM. The EBE has a degradation rate equal to $\Td$ 
		and we limit the functionality of the RM and IM by allowing only the maintenance action to perform cleaning. We also need the corresponding DELAY modules generating the degradation rates, $\Td$ and the maintenance rates $\Tcln,\Tin,\Trp$.
		The resulting CTMC is obtained by performing a parallel composition of the components $C_{all} = C_{EBE}||$ $C_{\Td}||C_{RM}||C_{IM}||C_{\Tcln}$ $||C_{\Tin}||C_{\Trp}.$ 
		\noindent The resulting state space is then $S_{all}= S_{EBE} \times S_{\Td} \times S_{RM} \times S_{IM} \times S_{\Tcln} \times S_{\Tin} \times S_{\Trp}  $. 
		The synchronisation between the different components is shown in Figure~\ref{fig:Synch} and proceeds as follows: 
		\begin{enumerate}
			\item All the DELAY modules (except $\Tcln$) start at the same time using the \texttt{trigger} transition label.
			\item When the extended DELAY module generating the $\Td$ time delay elapses, the corresponding EBE moves to the next state through synchronisation with the transition label \texttt{degrade}$_N$.
			\item The clock signals $\Trp,\Tin$ represent periodic maintenance and inspection actions and when the deterministic delay is reached, through synchronisation with the transition label \texttt{check\_clean} or the  $\texttt{inspect}$, the RM or IM modules are triggered (cf. Fig. \ref{fig:RM} and \ref{fig:IM}). If RM triggers a maintenance action, the DELAY representing $\Tcln$ is triggered using the synchronisation labels \texttt{trigger\_clean}. Once the deterministic delay $\Tcln$ elapses, the EBE, the extended DELAY module representing $\Td$ (where the \texttt{reset} transition label within the extended DELAY module is replaced with \texttt{perform\_clean}) and the IM are reset using the transition label \texttt{perform\_clean}.
		\end{enumerate}
		
		\begin{remark}\label{remark:ss}
			One should note that performing synchronisation results in a large state space, which is a function of the number of states used to approximate the deterministic delays. In order to counteract this effect we propose an abstraction framework in Subsection~\ref{subsec:FMTFrame:Dec}.
		\end{remark}

	
	\subsection{Metrics}
	\label{subsec:FMTFrame:Met}
	We use PRISM to compute the metrics of the model described in Subsection \ref{subsec:Prelim:FMTFrame}. The metrics can be expressed using the extended Continuous Stochastic Logic (CSL) as follows:
	\begin{enumerate}
		\item \textit{Reliability} :
		This can be expressed as the complement of the probability of failure over the time $T$, $1-\po_{=?}[\fo^{\le T} failed ]$.
		\item \textit{Availability}:
		This can be expressed 
		as $\ro_{=?}[C^{\le T} ]/T$, which corresponds to the cumulative reward of the total time spent in states labelled with \textit{okay} and \textit{thresh} during the time  $T$.
		\item \textit{Expected cost}:		
		This can be expressed using 
		$\ro_{=?}[C^{\le T}]$, which corresponds to the cumulative reward of the total costs (operational, maintenance and failure) within the time $T$.
		\item \textit{Expected number of failure}:		
		This can be expressed using 
		\noindent $\ro_{=?}[C^{\le T} ]$, which corresponds to the cumulative transition reward that counts the number of times the top event enters the \emph{failed state} within the time $T$.
		
	\end{enumerate}
	
	\subsection{Decomposition of FMTs}
	\label{subsec:FMTFrame:Dec}
\begin{figure*}[ht!]
		\psfragscanon
		\includegraphics[width=\textwidth]{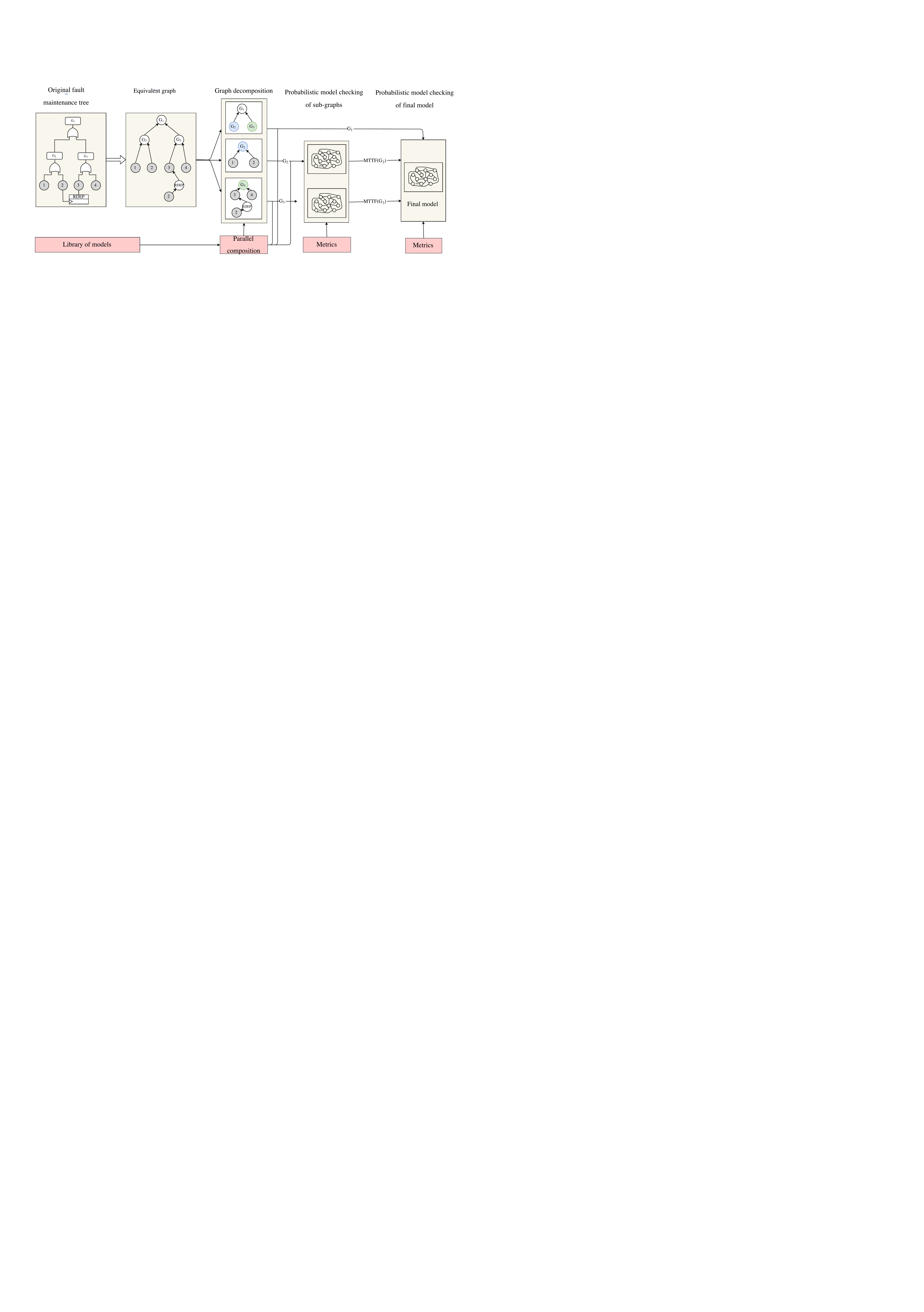}
		\caption{{Overall developed framework for decomposition of FMTs into the equivalent abstract CTMCs.}}
		\label{fig:Overall}
	\end{figure*}

	The use of CTMC and deterministic time delays results in a large state space  for modelling the whole FMT (cf. Remark~\ref{remark:ss}). We therefore propose an approach that decomposes the large FMT into an equivalent abstract CTMC that can be analysed using PRISM. The process involves two transformation steps. First we convert the FMT into an equivalent directed acyclic graph (DAG) and split this graph into a set of smaller sub-graphs. Second, we transform each sub-graph into an equivalent CTMC by making use of the developed FMT components semantics (cf. Subsec.~\ref{subsubsec:FMTFrame:Modelling:SemanticsElements}), and performing parallel composition of the individual FMT components based on the underlying structure of the sub-graph. The smaller sub-graphs are then sequentially composed to generate the higher level abstract FMT. Figure \ref{fig:Overall} depicts a high-level diagram of the decomposition procedure.
	
	\paragraph{\textbf{Conversion of the original FMT to the equivalent graph}}
	The FMT is a DAG (cf. Subsection~\ref{subsec:FMTFrame:Modelling}) and in this framework we need to apply a transformation to the DAG in the presence of an RDEP gate, such that we can perform the decomposition. The RDEP causes an acceleration of events on dependent children nodes when the input node fails. In order to capture this feature in a DAG,  we need to duplicate the input node such that it is connected directly to the RDEP vertex. This allows us to capture when the failure of the input occurs and the corresponding acceleration of the the children.  This is reasonable as the same RM and IM are used irrespective of the underlying FMT structure.	
	%
	\paragraph{\textbf{Graph decomposition}}
	We define modules within the DAG as sub-trees composed of at least two events which have no inputs from the rest of the tree and no outputs to the rest except from its output event~\cite{Li20151400}.	
	We can divide the graph into multiple partitions based on the number of modules making up the DAG.
	We define the following notations to ease the description of the algorithm:
	\begin{itemize}
		\item $V_o$ indicates whether the node is the top node of the DAG. 
		\item $V_g$ indicates the node where the graph split is performed.	
		\item Modules correspond to sub-graphs in DAG.
	\end{itemize} 
	We set $V_o$ when we construct the DAG from the FMT and then proceed with executing Algorithm \ref{alg:split}. We first identify all the sub-graphs within the whole DAG and label all the top nodes of each sub-graph $i$ as $V_{Ti}$. We loop through each sub-graph and its immediate child (the sub-graph at the immediate lower level) and at the point where the sub-graph and child are connected, the two graphs are split and a new node $V_g$ is introduced. Thus, executing Algorithm \ref{alg:split} results in a set of sub-graphs linked together by the labelled nodes $V_g$. For each of the lower-level sub-graphs, we now proceed to compute the mean time to failure (MTTF). This will serve as an input to the higher-level sub-graphs, such that metrics for the abstract equivalent CTMC can be computed.
	\begin{algorithm}
		\caption{\small{DAG decomposition algorithm}}
		\label{alg:split}
		\DontPrintSemicolon
		\LinesNumbered
		\SetKwInOut{Input}{input}\SetKwInOut{Output}{output}
		\label{alg:ADP_Nat1}
		\Input{ DAG $G=(V,E)$ }
		\Output{Set of sub-graphs with one of the end nodes labelled as $V_g$.}
		Identify sub-graphs using `depth-first' traversal \;
		Label all top nodes of each sub-graph $i$ as $V_{T_i}$ \;
		\ForAll{select the top node of every sub-graph and the child defined at the immediate lower level }{	
			\If{label $V_T$ already found in one of the leaf nodes of the sub-graph}{Split sub-graph \;
				Insert new node $V_{g}$ which will be used as input from connected sub-graph }
		}
	\end{algorithm} 

\begin{figure}[h!]
		\centering
		\includegraphics[width=\textwidth]{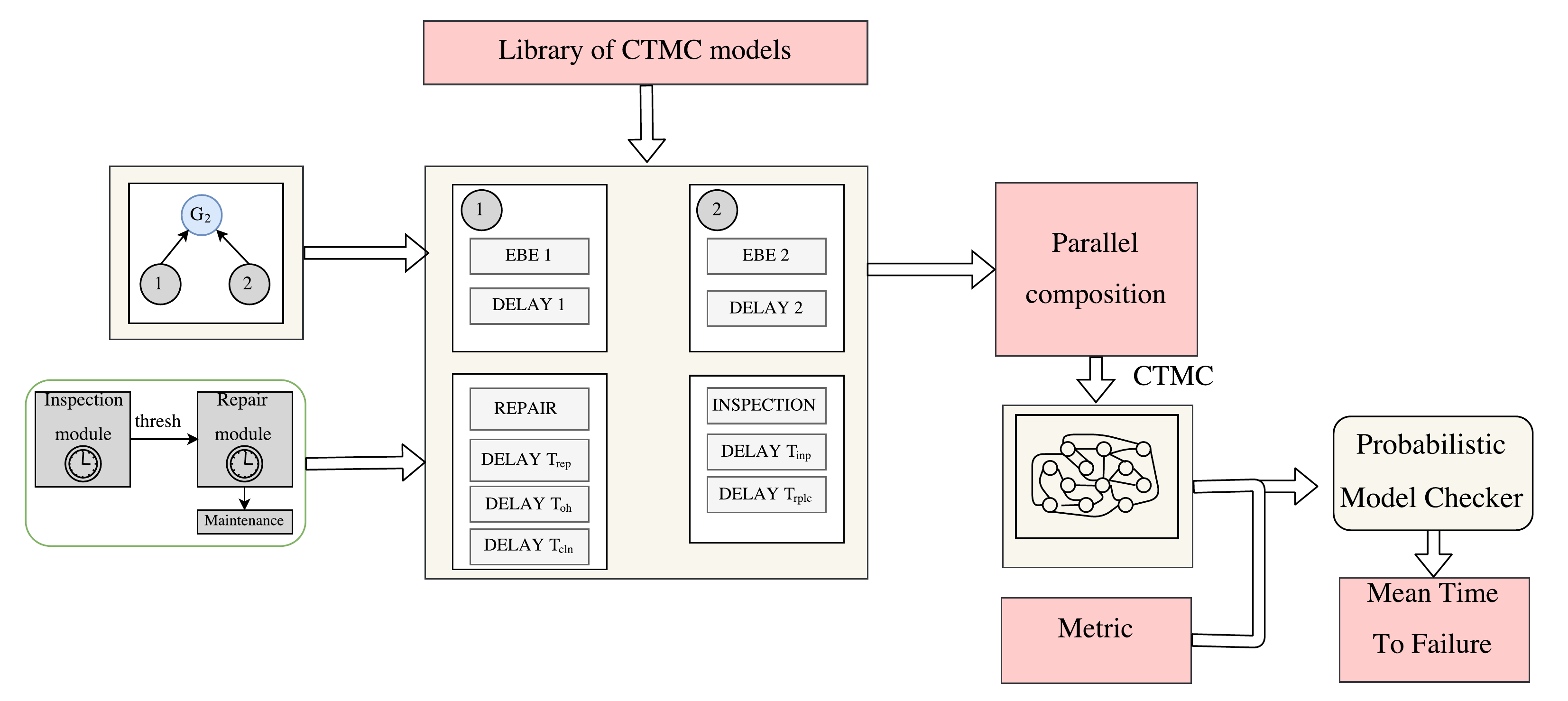}
		\caption{PMC of sub-graphs.}
		\label{fig:conv}
	\end{figure}

	\paragraph{\textbf{PMC of sub-graphs}}
	We start from the bottom level sub-graphs and perform the conversion to CTMC using the formal models presented in Subsection~\ref{subsubsec:FMTFrame:Modelling:SemanticsElements}. The formal models have been built into a library of PRISM modules and based on the underlying components and structure making up the sub-graph, the corresponding individual formal models are converted into the sub-graph's equivalent CTMC by performing parallel composition (cf.  Subsec. \ref{subsubsec:FMTFrame:Modelling:Semantics}). For each sub-graph, we compute the probability of failure $D_e(T)$ at time $T$
	, from which we calculate the MTTF~\cite{ruijters2015fault} using
	\begin{equation*}
	\mathit{MTTF} = \frac{\ln(1-D_e(T))}{-T}.
	\end{equation*}
	The MTTF serves as the input to the higher level sub-graph at time $T$. The new node in the higher-level sub-graph, now degrades with the new time delay $\Td = \mathit{MTTF}$, which is fed into the corresponding DELAY component. This process is repeated for all the different sub-graphs until the top level node $V_o$ is reached. {Figure~\ref{fig:conv} depicts the steps needed to perform PMC for one of the sub-graphs. }
	\paragraph{\textbf{PMC of final equivalent abstract CTMC}}
	On reaching the top level node $V_o$, we compute the metrics for the equivalent abstract CTMC for a specific time horizon $T$. For different horizons, the previous step of computing the MTTF for the underlying lower level sub-graphs needs to be repeated. 
	Using this technique, we can formally verify larger FMTs, while using less memory and computational time due to the significantly smaller state space  of the underlying CTMCs.
	\begin{figure}[h!]
	\centering
	\includegraphics[width=0.6\textwidth]{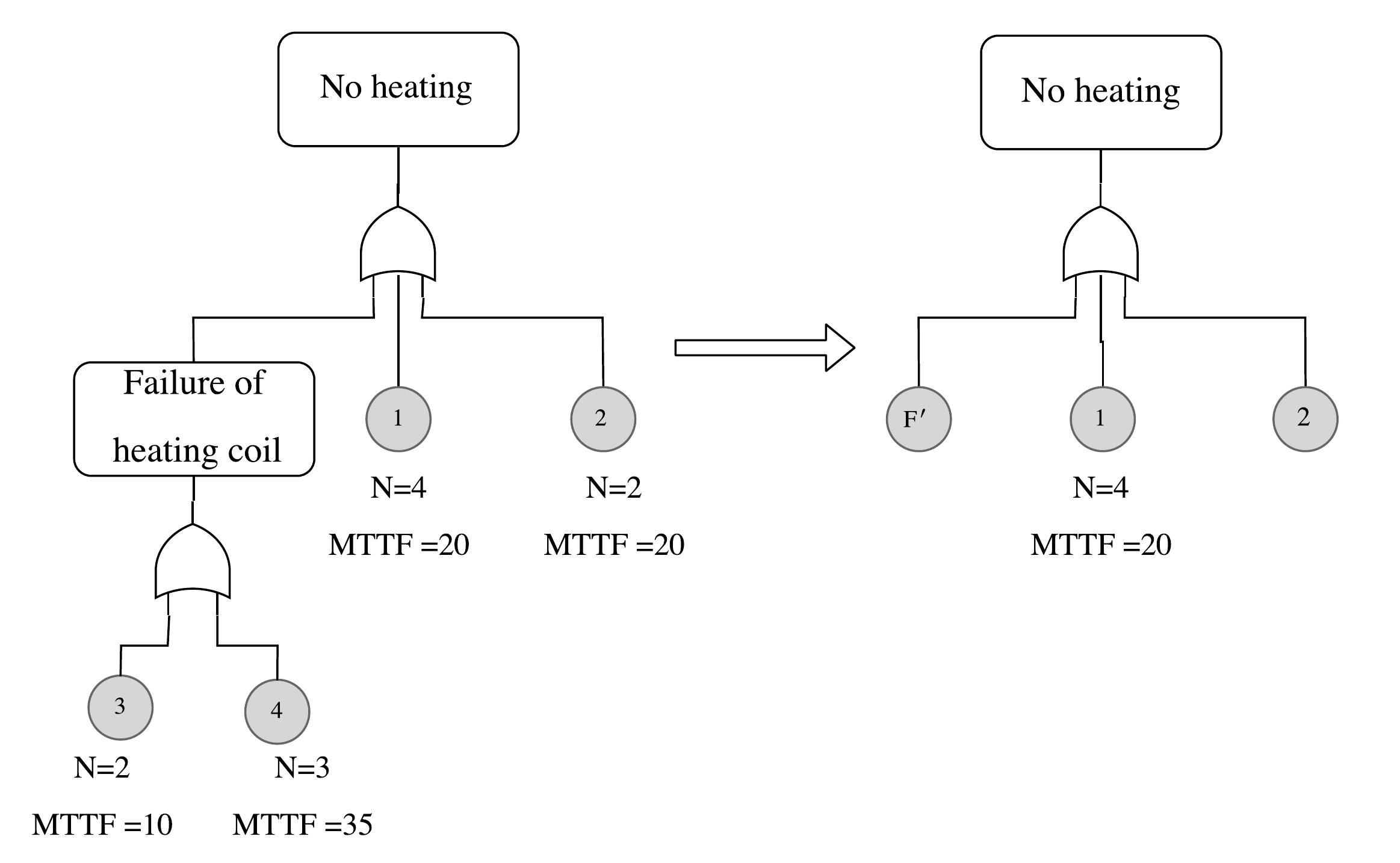}
	\caption{{The original FMT and the abstract FMT corresponding to the equivalent abstract CTMC generated by the developed framework. The MTTF for the F' is computed based on the probability of failure of the heating coil.}}
	\label{fig:IncVer}
	\end{figure}
	Next, we proceed with an illustrative example comparing the process of directly modelling the large FMT using CTMCs versus the de-compositional modelling procedure. Figure \ref{fig:IncVer} presents the FMT composed of two modules and the corresponding abstracted FMT. The abstract FMT is a pictorial representation of the model represented by the equivalent abstract CTMC obtained using the developed decomposition framework (cf. Fig. \ref{fig:Overall}). 	
	For both the large FMT and the equivalent abstract FMT a comparison between the total number of states for the resulting CTMC models, the total time to compute the reliability metric and the resulting reliability metric is performed. All computations are run on an 2.3 GHz Intel Core i5 processor  with 8 GB of RAM and the resulting statistics are listed in Table \ref{tab:HVAC_SizeEg}. The original FMT has a state space  with 193543 states, while the equivalent abstract CTMC has a state space  with 63937 states. This corresponds to a $67\%$ reduction in the state space  size. The total time to compute the reliability metric is a function of the final time horizon and a maximal $73\%$ reduction in computation time is achieved. Accuracy in the reliability metric of the abstract model is a function of the time horizon and the number of states used to approximate the deterministic delay representing the computed MTTF. The larger the number of states the more accurate the representation of the MTTF, but this comes at a cost on the size of the underlying CTMC model. In our case, $N=4$ is chosen. 
	The accuracy of the reliability metric computed by the abstract FMT results in a maximal reduction of $0.61\%$.

	\begin{table}[ht!]
		\centering
		\resizebox{0.7\textwidth}{!}{
		\begin{tabular}{ccccccc} 
			\hline
			{\textbf{Time}}& \multicolumn{2}{c}{\textbf{Original FMT}}  &  \multicolumn{4}{c}{\textbf{Abstracted FMT}} \\ 
			\textbf{Horizon}
			&{Time to compute} 		  & Reliability
			& \multicolumn{2}{c}{Time to compute}& Total  & Reliability\\    		
			&{metric} &             
			& MTTF   & {metric} & Time &\\
			(years) & (mins)   &      & (mins) & (mins) & (mins) & 			\\ \hline \hline
			5 				
			&\textbf{0.727}   & 0.9842      
			& 0.142   & 0.181  & \textbf{0.223}     & 0.9842\\ 
			10				
			&\textbf{1.406}   & 0.8761     
			& 0.219   & 0.309  & \textbf{0.528 }   & 0.8769\\
			15				
			&\textbf{2.489}   & 0.3290      
			& 0.292   & 0.622 & \textbf{0.914 }   & 0.3270 \\ \hline
		\end{tabular}}
		\caption{{Comparison between the original large FMT and the abstracted FMT.}}
		\label{tab:HVAC_SizeEg}
	\end{table}	
	\section{Case study}
	\label{sec:CaseStudy}
\begin{figure}[ht!]
		\centering
		\includegraphics[width=0.78\textwidth]{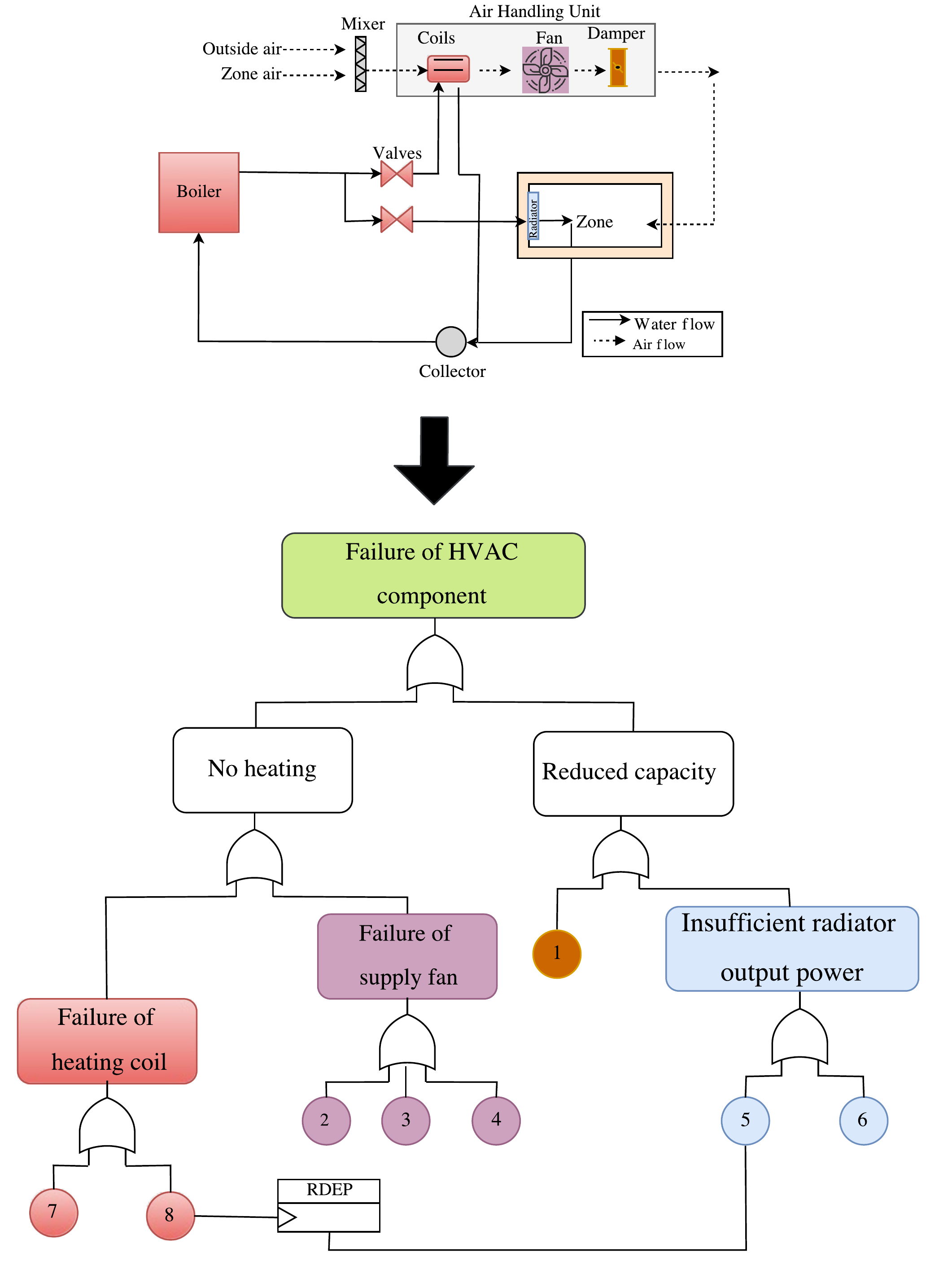}
		\caption{{FMT for failure in HVAC system with leaves represented using EBE (associated RM and IM not shown in figure). The EBE are labelled to correspond to the component failure they represent using the fault index presented in Table~\ref{tab:HVAC_FM}. The EBE and intermediate events are colour coded such that they correspond to the different HVAC components thus showing how the propagation of faults in the HVAC is reflected within the FMT.}}
		\label{fig:FMT}
	\end{figure}

	We apply the FMT framework to a Heating, Ventilation and Air-conditioning (HVAC) system used to regulate a building's internal environment (cf. Sec.~\ref{sec:hvac}).
	Based on this HVAC system we construct the corresponding FMT shown in Figure \ref{fig:FMT}. The FMT structure follows the structure of the underlying HVAC system, as can be seen from the colour shading used in Figure \ref{fig:FMT}. The leaves of the tree are EBE with discrete degradation rates computed using Table \ref{tab:HVAC_FM}, approximated by the Erlang distribution where $N$ is the number of degradation phases ($k=N$ for the Erlang distribution) and MTTF is the expected time to failure with $MTTF = 1/\lambda$ (cf. Remark~\ref{remark:Er}). We choose an acceleration factor $\gamma = 2$ for the RDEP gate. The system is periodically cleaned every $\Trp$ months 
	and a major overhaul with a complete replacement of all components is carried out once every $\Toh$ years. 
	Inspections are performed every $\Tin$ months and 
	return the components back to the previous state, corresponding to a cleaning action. 
	The total time to perform a cleaning action is 1 day ($\Tcln= 1\;day$), while performing a total replacement of components takes 7 days ($\Trpl = 7\;days$). The time timing signals $\{\Trp, \Toh,\Tin,\Tcln,\Trpl\}$ are all approximated using the Erlang distribution with $N = 3$. All maintenance actions are performed simultaneously on all components. 
	\begin{table}[ht!]
		\centering
		\begin{tabular}{*5c} 
			\hline
			\textbf{Fault Index } & \textbf{Failure Mode} & \textbf{N} & \textbf{MTTF}\\ 
			  & 					    &  &\textbf{(years)} \\ \hline \hline
			1 & Broken AHU Damper & 4 & 20 \\ 
			2 & Fan motor failure & 3 & 35 \\ 
			3 & Obstructed supply fan& 4 & 31 \\
			4 & Fan bearing failure &6 &17 \\ 
			5 & Radiator failure & 4 & 25 \\ 
			6 & Radiator stuck valve & 2 & 10 \\  
			7 & Heater stuck valve & 2&10 \\ 
			8 & Failure in heat pump &4&20 \\ \hline
		\end{tabular}
		\caption{{Extended Basic events in FMT with associated degradation rates (N, MTTF) obtained from  ~\cite{handbook1996hvac,Khan2003561}.}}
		\label{tab:HVAC_FM}
	\end{table}  
	
	\subsection{Quantitative results}
	\label{subsec:CaseStudy:Res}
	In the following subsections, we employ the developed framework (cf. Subsec. \ref{subsec:FMTFrame:Dec}) to the FMT representing the failure of the HVAC system (cf. Fig. \ref{fig:FMT}) and perform three different experiments. 
	We first demonstrate the use of the developed framework by converting the FMT for the HVAC set-up into an abstract CTMC.
	For this abstract CTMC we compute the metrics (cf. Sec. \ref{subsec:FMTFrame:Met}) using probabilistic model checking to show the type of analysis that can be performed using the set-up. Next, we perform a comparison between different maintenance strategies applied to the same FMT. This allows the user to deduce the optimal strategy for the set-up. Last, we construct a FMT which does not employ the repair and inspection module and compare it with the original FMT (includes the maintenance modules) to further highlight the advantage of incorporating maintenance.
	
	\paragraph{\textbf{Applying the framework to HVAC set-up}}
	\label{subsec:CaseStudy:Res:1}
	
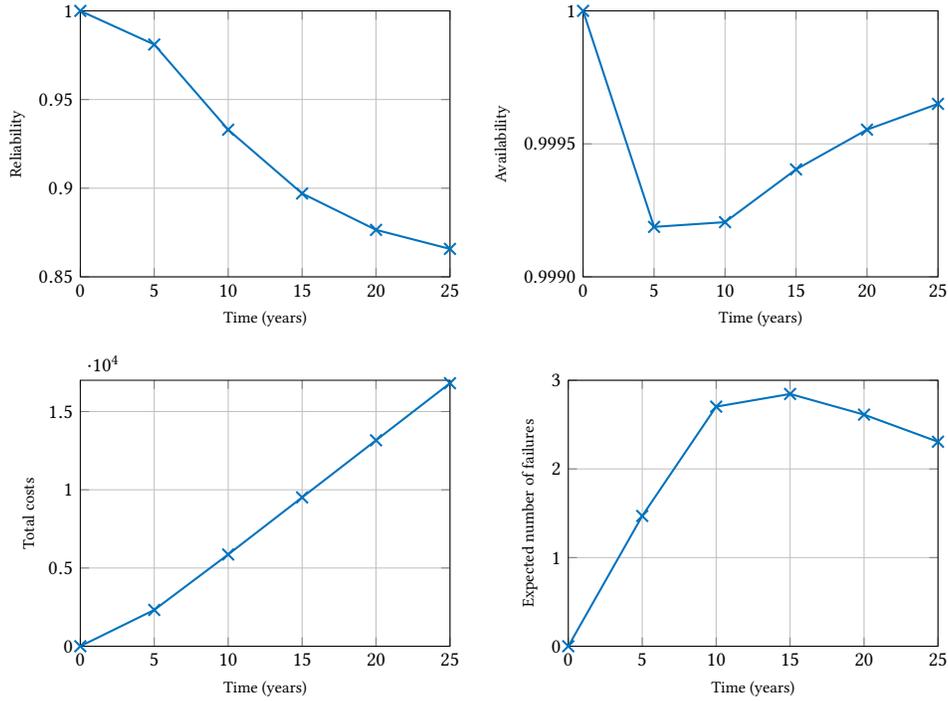
\begin{figure}[ht!]
		\centering
		\psfragscanon
		\resizebox{0.8\textwidth}{!}{
%
%
\definecolor{mycolor1}{rgb}{0.00000,0.44700,0.74100}%
\definecolor{mycolor2}{rgb}{0.85000,0.32500,0.09800}%
\definecolor{mycolor3}{rgb}{0.92900,0.69400,0.12500}%
\definecolor{mycolor4}{rgb}{0.49400,0.18400,0.55600}%
\definecolor{mycolor5}{rgb}{0.46600,0.67400,0.18800}%
\definecolor{mycolor6}{rgb}{0.30100,0.74500,0.93300}%
\definecolor{mycolor7}{rgb}{0.63500,0.07800,0.18400}%
\begin{tikzpicture}

\begin{axis}[%
width=2.5in,
height=1.8in,
at={(0in,0in)},
scale only axis,
xmin=0,
xmax=25,
xmajorgrids,
xlabel={\footnotesize Time (years)},
ymin=0.85,
ymax=1,
ymajorgrids,
ylabel={\footnotesize Reliability},
axis background/.style={fill=white},
title style={font=\bfseries},
xtick={0,5,10,15,20,25},
xticklabels={0,5,10,15,20,25},
legend style={legend cell align=left,align=left,draw=white!15!black}
]
\addplot [color=mycolor1,mark=x, solid,line width=1pt,mark options={scale=2}]
  table[row sep=crcr]{%
	0	1\\
5   0.9810270134546714  \\
10  0.9329803505882919 \\ 
15  0.8970186674104699 \\
20  0.8764872878480361 \\ 
25  0.8657625484657094 \\
};
\end{axis}

\begin{axis}[%
width=2.4in,
height=1.8in,
at={(3.4in,0in)},
scale only axis,
xmin=0,
xmax=25,
xmajorgrids,
xlabel={\footnotesize Time (years)},
ymin=0.9990,
ymax=1,
ymajorgrids,
ylabel={\footnotesize Availability},
axis background/.style={fill=white},
title style={font=\bfseries},
ytick={1,0.9995,0.9990},
yticklabels={1,0.9995,{0.9990}},
xtick={0,5,10,15,20,25},
xticklabels={0,5,10,15,20,25},
legend style={legend cell align=left,align=left,draw=white!15!black}
]
\addplot [color=mycolor1,mark=x, solid,line width=1pt,mark options={scale=2}]
table[row sep=crcr]{%
	0	1\\
 5  0.9991877869304348 \\
10  0.9992056474381137  \\
15  0.9994039567025734  \\
20  0.9995533878059931  \\
25  0.9996503214008801  \\
};
\end{axis}

\begin{axis}[%
width=2.5in,
height=1.8in,
at={(0in,-2.5in)},
scale only axis,
xmin=0,
xmax=25,
xmajorgrids,
xlabel={\footnotesize Time (years)},
ymin=0,
ymax=17000,
ymajorgrids,
ylabel={\footnotesize Total costs},
axis background/.style={fill=white},
title style={font=\bfseries},
xtick={0,5,10,15,20,25},
xticklabels={0,5,10,15,20,25},
legend style={legend cell align=left,align=left,draw=white!15!black}
]
\addplot [color=mycolor1,mark=x, solid,line width=1pt,mark options={scale=2}]
table[row sep=crcr]{%
0 0\\
5 2320.186754272556\\
10 5865.051409984242 \\
15 9514.384428850382 \\
20 13163.522992007418\\
25  16812.49248394761\\
};
\end{axis}

\begin{axis}[%
width=2.5in,
height=1.8in,
at={(3.3in,-2.5in)},
scale only axis,
xmin=0,
xmax=25,
xmajorgrids,
xlabel={\footnotesize Time (years)},
ymin=0,
ymax=3,
ymajorgrids,
ylabel={\footnotesize Expected number of failures},
axis background/.style={fill=white},
title style={font=\bfseries},
xtick={0,5,10,15,20,25},
xticklabels={0,5,10,15,20,25},
legend style={legend cell align=left,align=left,draw=white!15!black}
]
\addplot [color=mycolor1,mark=x,solid,line width=1pt,mark options={scale=2}]
table[row sep=crcr]{%
	0	0\\
 5   1.4696429270974607 \\
10  2.7025212368046048\\
15  2.8455197057855277\\
20  2.6125965637888995\\
25  2.3065375134174615\\
};
\end{axis}

\end{tikzpicture}%
}
		\caption{{Reliability, availability, total costs and expected number of failures of HVAC over time horizon $N_r= \{0,5,10,15,20,25\}$.}}
		\label{fig:Rel_Avail}
	\end{figure}

	We convert the FMT representing the failure of the HVAC system  into the equivalent abstract CTMC and perform probabilistic model checking over six time horizons $N_r= \{0,5,10,15,20,$ $25\}$ years with the maintenance policy consisting of periodic cleaning every {$\Trp = 2 $ years} and inspections 
	every { $\Tin = 1$ year}. No replacement actions are considered.
	For this set-up, all the metrics corresponding to the reliability, availability, total costs (maintenance, inspection and operational costs) and the total expected number of failures of the HVAC systems over the time horizon are computed and are shown in Figure \ref{fig:Rel_Avail}. The total maintenance cost to perform a clean is 100 [GBP], while an inspection cost 50 [GBP]. 
	The maximal time taken to compute a metric using the abstract FMT is 1.47 minutes. It is deduced that the reliability reduces over time. The availability is seen to be nearly constant, while the expected number of failures increases until it reaches a steady state value. This shows that there is a saturation in the number of maintenance actions which one can perform before the system no longer achieves higher performance in reliability and availability.  One can further note that, as expected, the  maintenance costs increases linearly with time.
	
	\paragraph{\textbf{Comparison between different maintenance strategies}}
	\label{subsec:CaseStudy:Res:2}
	
	In this second experiment, we compare all the metrics (reliability, availability, total costs and expected number of failures) over the time horizon $N_r= \{0,5,10,15,20,25\}$ years when considering different maintenance strategies, such that we can identify the optimal maintenance strategy that minimises cost and achieves the best trade-off in HVAC performance (i.e. with minimal expected number of failures and high reliability and availability). We consider five different maintenance strategies which are listed in Table \ref{tab:MaintStrat}. 
	\begin{table}[ht!]
		\centering
		\begin{tabular}{*5c} 
			\hline
			\textbf{Strategy index} & {$\Trp$} & $\Toh$ & $\Tin$\\ \hline \hline
			$M_0$ & 2 years    & -  & 1 year \\
			$M_1$ & 5 years    & -  & 2 years  \\ 
			$M_2$ & 2 years    & 5 years & - \\ 
			$M_3$ & 2 years    & 10 years & 1 year \\
			$M_4$ & 2 years   & 20 years & 6 months \\ \hline \hline
		\end{tabular}
		\caption{{Implemented maintenance strategies.}}
		\label{tab:MaintStrat}
	\end{table}	

\begin{figure}[h!]
		\psfragscanon \centering
		\begin{tikzpicture}
		\begin{axis}[%
		hide axis,
		xmin=10,
		xmax=50,
		ymin=0,
		ymax=0.4,
		legend style={at={(0,0)},anchor=north,legend columns=6,legend cell align=left,align=left,draw=white!15!black,legend cell align=left}
		]
		\addlegendimage{color=mycolor1,solid,line width=1pt,mark=x,mark options={scale=2}}
		\addlegendentry{ $M_0$};
		\addlegendimage{color=mycolor2,dotted,line width=1pt,mark=+,mark options={scale=2}}
		\addlegendentry{ $M_1$};
		\addlegendimage{color=mycolor3,dashed,line width=1pt,mark=triangle,mark options={scale=2}}
		\addlegendentry{ $M_2$};
		\addlegendimage{color=mycolor4,dash dot,line width=1pt,mark=diamond,mark options={scale=2}}
		\addlegendentry{ $M_3$};
		\addlegendimage{color=mycolor5,dashed,line width=1pt,mark=square,mark options={scale=2}}
		\addlegendentry{ $M_4$};
		\end{axis}
		\end{tikzpicture}%
		\\
		\centering
		\resizebox{0.8\textwidth}{!}{
%
%
\definecolor{mycolor1}{rgb}{0.00000,0.44700,0.74100}%
\definecolor{mycolor2}{rgb}{0.85000,0.32500,0.09800}%
\definecolor{mycolor3}{rgb}{0.92900,0.69400,0.12500}%
\definecolor{mycolor4}{rgb}{0.49400,0.18400,0.55600}%
\definecolor{mycolor5}{rgb}{0.46600,0.67400,0.18800}%
\definecolor{mycolor6}{rgb}{0.30100,0.74500,0.93300}%
\definecolor{mycolor7}{rgb}{0.63500,0.07800,0.18400}%

\begin{tikzpicture}
\begin{axis}[%
width=2.5in,
height=1.9in,
at={(0in,0in)},
scale only axis,
xmin=0,
xmax=25,
xmajorgrids,
xlabel={\footnotesize Time (years)},
ymin=0.4,
ymax=1,
ymajorgrids,
ylabel={\footnotesize Reliability},
axis background/.style={fill=white},
title style={font=\bfseries},
xtick={0,5,10,15,20,25},
xticklabels={0,5,10,15,20,25}
]
\addplot [color=mycolor1,mark=x, solid,line width=1pt,mark options={scale=2}]
table[row sep=crcr]{%
	0   1\\
5   0.9810270134546714 \\
10  0.9329803505882919\\
15  0.8970186674104699\\
20  0.8764872878480361\\
25  0.8657625484657094\\
};
\addplot [color=mycolor2,mark=+,mark=+,dotted,line width=1pt,mark options={scale=2}]
table[row sep=crcr]{%
	0   1\\
	5	0.9674939208705396\\
	10	0.8273667890815352\\
	15	0.6897378082156743\\
	20	0.6028696555345895\\
	25	0.5551807725736861\\
};
\addplot [color=mycolor3,mark=triangle,dashed,line width=1pt,mark options={scale=2}]
table[row sep=crcr]{%
	0   1\\
5   0.9823520585142311\\
10  0.9404652694646075\\
15  0.9098766789870915\\
20  0.8593944722338369\\
25  0.8835920889243212\\
};
\addplot [color=mycolor4,mark=diamond,dash dot,line width=1pt,mark options={scale=2}]
table[row sep=crcr]{%
	0   1\\
5  0.9812958171423642 \\
10  0.9347100484199586\\
15  0.9001115714934607\\
20  0.8804331528627425\\
25  0.8701779503575927\\
};
\addplot [color=mycolor5,mark=square,dashed,line width=1pt,mark options={scale=2}]
table[row sep=crcr]{%
	0   1\\
 5  0.9810270134546776 \\
10 0.9329803505883203\\
15 0.8970186674104818\\
20 0.8764872878479744\\
25 0.8657625484657038\\
};
\end{axis}

\begin{axis}[%
width=2.4in,
height=1.9in,
at={(3.4in,0in)},
scale only axis,
xmin=0,
xmax=25,
xmajorgrids,
xlabel={\footnotesize Time (years)},
ymin=0.99,
ymax=1,
ymajorgrids,
ylabel={\footnotesize Availability},
axis background/.style={fill=white},
title style={font=\bfseries},
xtick={0,5,10,15,20,25},
xticklabels={0,5,10,15,20,25},
ytick={1,0.995, 0.990},
yticklabels={1,0.995, 0.990}
]
\addplot [color=mycolor1,solid,mark=x, line width=1pt,mark options={scale=2}]
table[row sep=crcr]{%
	0 1\\
 5 0.9991877869304348 \\
10 0.9992056474381137 \\
15 0.9994039567025734 \\
20 0.9995533878059931 \\
25 0.9996503214008801 \\
};

\addplot [color=mycolor2,mark=+,dotted,line width=1pt,mark options={scale=2}]
table[row sep=crcr]{%
	0 1\\
5   0.9947178195772173 \\
10  0.9903207806288671 \\
15  0.9914612330659673 \\
20  0.9933847722789865 \\
25  0.9949830115793791  \\
};
\addplot [color=mycolor3,mark=triangle,dashed,line width=1pt,mark options={scale=2}]
table[row sep=crcr]{%
	0 1\\
 5  0.9999708462757406 \\
10  0.9999497275797525 \\
15  0.9999484219631631 \\
20  0.9999379342018608 \\
25  0.9999594408415446 \\
};
\addplot [color=mycolor4,mark=diamond,dash dot,line width=1pt,mark options={scale=2}]
table[row sep=crcr]{%
	0 1\\
5   0.9992114814417806 \\
10  0.9992344957693821 \\
15  0.999424034451488  \\
20  0.9995650661285602\\
25  0.9992077926589134\\
};
\addplot [color=mycolor5,mark=square,densely dashed,line width=1pt,mark options={scale=2}]
table[row sep=crcr]{%
	0 1\\
5	0.9995184732407355 \\
10	0.9995095552918987\\
15	0.9996140645452434\\
20	0.9996944633498803\\
25	0.9997466811982463\\
};
\end{axis}
\begin{axis}[%
width=2.5in,
height=1.9in,
at={(0in,-2.5in)},
scale only axis,
xmin=0,
xmax=25,
xmajorgrids,
xlabel={\footnotesize Time (years)},
ymin=0,
ymax=30000,
ymajorgrids,
ylabel={\footnotesize Total cost},
axis background/.style={fill=white},
title style={font=\bfseries},
xtick={0,5,10,15,20,25},
xticklabels={0,5,10,15,20,25}
]
\addplot [color=mycolor1,mark=x, solid,line width=1pt,mark options={scale=2}]
  table[row sep=crcr]{%
0   0\\
5   2320.186754272556  \\
10  5865.051409984242  \\
15  9514.384428850382  \\
20  13163.522992007418 \\
25  16812.49248394761  \\
};
\addplot [color=mycolor2,mark=+,dotted,line width=1pt,mark options={scale=2}]
table[row sep=crcr]{%
0   0\\
5	1869.4284024210124\\
10	4317.036405879204 \\
15	7576.450290246769\\
20	11144.892299191399\\
25	14772.00641037058\\
};
\addplot [color=mycolor3,mark=triangle,dashed,line width=1pt,mark options={scale=2}]
table[row sep=crcr]{%
0   0\\
5   3542.3269985073694  \\
10  10147.31162485666   \\
15  16846.752326907106  \\
20  23523.513765141444  \\
25  30280.939988186947  \\
};
\addplot [color=mycolor4,mark=diamond,dash dot,line width=1pt,mark options={scale=2}]
table[row sep=crcr]{%
0   0\\
5   2652.321288346727 \\
10  7277.35253818163   \\
15  12036.578314185745 \\
20  16797.901328879565 \\
25  21527.6693752089   \\
};
\addplot [color=mycolor5,mark=square,densely dashed,line width=1pt,mark options={scale=2}]
table[row sep=crcr]{%
0	0\\
5	2339.302690055152 \\
10	5886.411157283551 \\
15	9535.818960953633\\
20	13184.828139076157 \\
25	16833.63385126899 \\
};
\end{axis}
\begin{axis}[%
width=2.5in,
height=1.9in,
at={(3.3in,-2.5in)},
scale only axis,
xmin=0,
xmax=25,
xmajorgrids,
xlabel={\footnotesize Time (years)},
ymin=0,
ymax=50,
ymajorgrids,
ylabel={\footnotesize Expected number of failures},
axis background/.style={fill=white},
title style={font=\bfseries},
xtick={0,5,10,15,20,25},
xticklabels={0,5,10,15,20,25}
]
\addplot [color=mycolor1,mark=x, solid,line width=1pt,mark options={scale=2}]
table[row sep=crcr]{%
0 0\\
5  1.4696429270974607\\
10 2.7025212368046048\\
15 2.8455197057855277\\
20 2.6125965637888995\\
25 2.3065375134174615\\
};
\addplot [color=mycolor2,mark=+,dotted,line width=1pt,mark options={scale=2}]
table[row sep=crcr]{%
	0 0\\
	5 9.639979271753349\\
	10 35.32915070511152\\
	15 46.71413860671433\\
	20 47.84152339009347\\
	25 44.808511179085926\\
};
\addplot [color=mycolor3,mark=triangle,dashed,line width=1pt,mark options={scale=2}]
table[row sep=crcr]{%
	0 0\\
	5 0.02640160631394073\\
	10 0.06434367605204\\
	15 0.08532871744312316\\
	20 0.1258853770917041\\
	25 0.09752019923233157\\
};
\addplot [color=mycolor4,mark=diamond,dash dot,line width=1pt,mark options={scale=2}]
table[row sep=crcr]{%
	0 0\\
		0 0\\
5  1.417902639824363\\
10 2.5675072694936705\\
15 2.6883956395842703\\
20 2.4609877668608897\\
25 2.168752150324141\\
};
\addplot [color=mycolor5,mark=square,densely dashed,line width=1pt,mark options={scale=2}]
table[row sep=crcr]{%
	0 0\\
	5 0.841200556465098\\
	10 1.5545049674463312\\
	15 1.6469917310482165\\
	20 1.521472522847438\\
	25 1.351301420839612\\
};
\end{axis}

\end{tikzpicture}%
}
		\caption{{Comparison between different number of maintenance strategies for an HVAC systems. }}
		\label{fig:NumFailures}
	\end{figure}
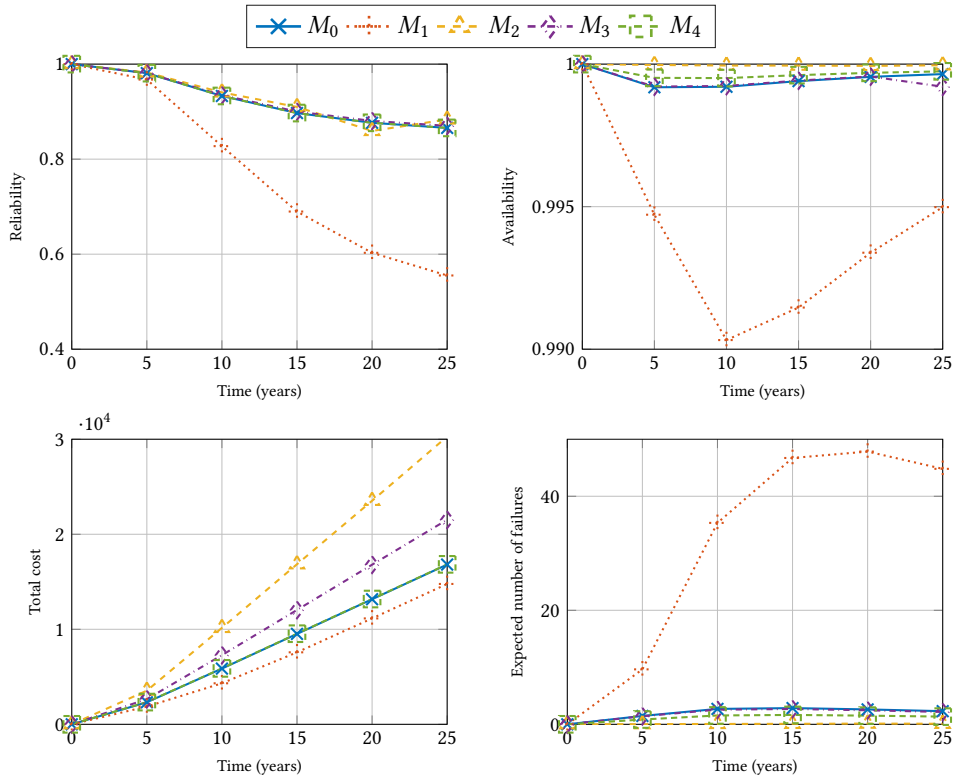

	We select strategies that have a different combination of repair, inspection and replacement strategies to highlight the effect the different maintenance actions have on the HVAC system's performance. Figure~\ref{fig:NumFailures} depicts the resulting metrics for the employed strategies.	

	We can deduce that the worst performing strategy is when cleaning actions are carried out every 5 years with inspection carried out bi-annually and no replacements (corresponding to strategy $M_1$). Strategies $M_2$ and $M_3$ have comparable high performance but with a  significant increase in the total costs due to the replacement action. We witness the highest costs using strategy $M_2$ due to the frequent replacement of the HVAC system. Comparing strategies $M_3$  and $M_4$ we can note that $M_3$ has fewer number of failures over the whole time horizon but this comes with higher total costs due to the replacements. Strategies $M_0$ and $M_4$ have similar performance with $M_0$ having a slightly lower availability and higher expected number of failures but with comparable maintenance costs.
	From this analysis, we can deduce that the optimal strategy which gives the best trade-off between costs and HVAC system's performance is strategy $M_0$ (i.e. with annual inspections, bi-annual cleaning and no replacements).  

	\paragraph{\textbf{Comparison between performing maintenance and no maintenance}}
 	Lastly, we compare the performance of the HVAC system without performing any maintenance actions vs the HVAC system with annual inspections, bi-annual cleaning and a major overhaul after 10 years. We employ the developed framework to represent the FMT of the HVAC system, first without incorporating the repair and inspection modules and then incorporating the repair and inspection modules with $\Tin = 1$ year, $\Trp=2$ years and $\Toh=10$ years.
 		\label{subsec:CaseStudy:Res:3}
 	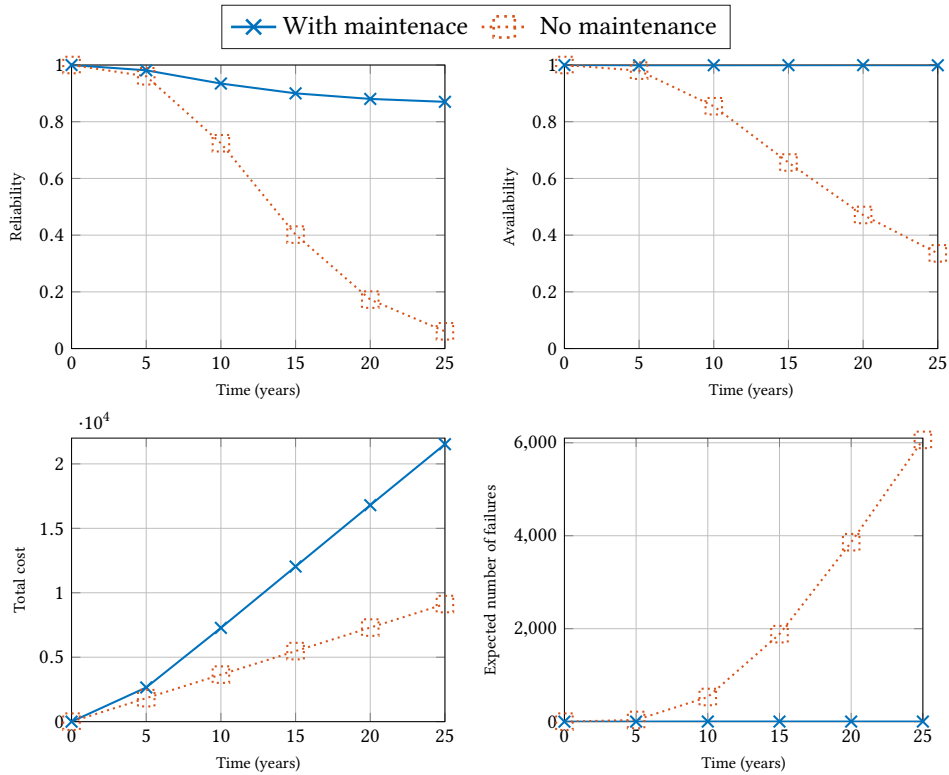
\begin{figure}[h!]
 		\psfragscanon \centering
 		\definecolor{mycolor1}{rgb}{0.00000,0.44700,0.74100}%
 		\begin{tikzpicture}
 		\begin{axis}[%
 		hide axis,
 		xmin=10,
 		xmax=50,
 		ymin=0,
 		ymax=0.4,
 		legend style={at={(0,0)},anchor=north,legend columns=6,legend cell align=left,align=left,draw=white!15!black,legend cell align=left}
 		]
 		\addlegendimage{color=mycolor1,solid,line width=1pt,mark=x,mark options={scale=2}}
 		\addlegendentry{ $\text{With maintenace }  $};
 		\addlegendimage{color=mycolor2,dotted,line width=1pt,mark=square,mark options={scale=2}}
 		\addlegendentry{ $\text{ No maintenance}$};
 		\end{axis}
 		\end{tikzpicture}%
 		\\
 		\centering
 		\resizebox{0.8\textwidth}{!}{
%
%
\definecolor{mycolor1}{rgb}{0.00000,0.44700,0.74100}%
\definecolor{mycolor2}{rgb}{0.85000,0.32500,0.09800}%
\definecolor{mycolor3}{rgb}{0.92900,0.69400,0.12500}%
\definecolor{mycolor4}{rgb}{0.49400,0.18400,0.55600}%
\definecolor{mycolor5}{rgb}{0.46600,0.67400,0.18800}%
\definecolor{mycolor6}{rgb}{0.30100,0.74500,0.93300}%
\definecolor{mycolor7}{rgb}{0.63500,0.07800,0.18400}%
%
%
%
\definecolor{mycolor1}{rgb}{0.00000,0.44700,0.74100}%
\definecolor{mycolor2}{rgb}{0.85000,0.32500,0.09800}%
\definecolor{mycolor3}{rgb}{0.92900,0.69400,0.12500}%
\definecolor{mycolor4}{rgb}{0.49400,0.18400,0.55600}%
\definecolor{mycolor5}{rgb}{0.46600,0.67400,0.18800}%
\definecolor{mycolor6}{rgb}{0.30100,0.74500,0.93300}%
\definecolor{mycolor7}{rgb}{0.63500,0.07800,0.18400}%

\begin{tikzpicture}
\begin{axis}[%
width=2.5in,
height=1.9in,
at={(0in,0in)},
scale only axis,
xmin=0,
xmax=25,
xmajorgrids,
xlabel={\footnotesize Time (years)},
ymin=0,
ymax=1,
ymajorgrids,
xtick={0,5,10,15,20,25},
ylabel={\footnotesize Reliability},
axis background/.style={fill=white},
title style={font=\bsfseries},
]
\addplot [color=mycolor1,mark=x, solid,line width=1pt,mark options={scale=2}]
table[row sep=crcr]{%
	0   1\\
	5	0.9812958171423642  \\
	10	0.9347100484199586\\
	15	0.9001115714934607\\
	20	0.8804331528627425\\
	25	0.8701779503575927 \\
};
\addplot [color=mycolor2,mark=square,dotted,line width=1pt,mark options={scale=2}]
table[row sep=crcr]{%
	0   1\\
	5	0.9603616221080067  \\
	10	0.7241387284378815\\
	15	0.4018712464478299\\
	20	0.17271034897337234\\
	25 0.060945009013046336\\
};
\end{axis}

\begin{axis}[%
width=2.5in,
height=1.9in,
at={(3.3in,0in)},
scale only axis,
xmin=0,
xmax=25,
xmajorgrids,
xtick={0,5,10,15,20,25},
xlabel={\footnotesize Time (years)},
ymin=0,
ymax=1,
ymajorgrids,
ylabel={\footnotesize Availability},
axis background/.style={fill=white},
title style={font=\bfseries}
]
\addplot [color=mycolor1,mark=x, solid,line width=1pt,mark options={scale=2}]
table[row sep=crcr]{%
	0 1\\
	5 0.9992114814417806\\
	10 0.9992344957693821 \\
	15 0.999424034451488 \\
	20 0.9995650661285602\\
	25 0.9992077926589134 \\
};
\addplot [color=mycolor2,mark=square,dotted,line width=1pt,mark options={scale=2}]
table[row sep=crcr]{%
	0 1\\
	5  0.9800581426845268\\
	10 0.8547019932341636\\
	15 0.6561765772150384\\
	20 0.47115672635005673\\
	25 0.33571530257668597 \\
};
\end{axis}
\begin{axis}[%
width=2.5in,
height=1.9in,
at={(0in,-2.5in)},
scale only axis,
xmin=0,
xmax=25,
xmajorgrids,
xlabel={\footnotesize Time (years)},
ymin=0,
ymax=22000,
ymajorgrids,
xtick={0,5,10,15,20,25},
ylabel={\footnotesize Total cost},
axis background/.style={fill=white},
title style={font=\bfseries},
]
\addplot [color=mycolor1,mark=x, solid,line width=1pt,mark options={scale=2}]
table[row sep=crcr]{%
	0   0\\
	5	2652.321288346727\\
	10	7277.35253818163 \\
	15	12036.578314185745\\
	20	16797.901328879565\\
	25	21527.6693752089\\
};
\addplot [color=mycolor2,mark=square,dotted,line width=1pt,mark options={scale=2}]
table[row sep=crcr]{%
	0   0\\
	5	1825\\
	10	3650\\
	15	5475\\
	20	7300\\
	25	9125\\
};
\end{axis}

\begin{axis}[%
width=2.4in,
height=1.9in,
at={(3.3in,-2.5in)},
scale only axis,
xmin=0,
xmax=25,
xmajorgrids,
xlabel={\footnotesize Time (years)},
ymin=0,
ymax=6100,
ymajorgrids,
xtick={0,5,10,15,20,25},
ylabel={\footnotesize Expected number of failures},
axis background/.style={fill=white},
title style={font=\bfseries},
]
\addplot [color=mycolor1,mark=x, solid,line width=1pt,mark options={scale=2}]
table[row sep=crcr]{%
	0 0\\
	5 1.417902639824363\\
	10 2.5675072694936705\\
	15 2.6883956395842703\\
	20 2.4609877668608897\\
	25 2.168752150324141\\
};
\addplot [color=mycolor2,mark=square,dotted,line width=1pt,mark options={scale=2}]
table[row sep=crcr]{%
	0 0\\
	5 36.39388960131779\\
	10 530.3377246940374\\
	15 1882.4332397520604\\
	20 3860.5558976491684\\
	25 6061.597863992404\\
};
\end{axis}

\end{tikzpicture}%
}
 		\caption{Comparison between incorporating the maintenance modules vs performing no maintenance. }
 		\label{fig:CompRMIM}
 	\end{figure}
 	The obtained results, depicted in Figure~\ref{fig:CompRMIM}, highlight the importance of maintenance and how appropriate maintenance strategies are required in order to maintain a reliable and available HVAC. When no maintenance is performed, both the reliability and availability of the HVAC system are gradually reduced, while the expected number of failures increases, as the components are degrading with time. This is in contrast to when maintenance is performed where high performance values of reliability and availability are achieved and the expected number of failures are low, throughout the whole time horizon.  One should note, that this comes at a price, where the total costs increase when maintenance is applied. Consequently, this further highlights the need to perform an analysis to deduce the optimal maintenance strategy which gives the best trade-off between costs, reliability, availability and the expected number of failures. 

	\section{Conclusion and Future Works}
	\label{sec:Conc}
	The paper presents a methodology for applying probabilistic model checking to FMTs. We model FMTs using CTMCs which simplify the transformation of FMT into formal models that can be analysed using PRISM. We further present a novel technique for abstracting the equivalent CTMC model. The novel decomposition procedure tackles the issue of state space  explosion and results in a significant reduction in both the state space  size and the total time required to compute metrics.  The framework is applied to an HVAC system and a set of different experiments to demonstrate the use of the developed framework and to highlight (i) the importance of performing maintenance and (ii) the effect of applying different maintenance strategies has been presented.  The presented framework can be further enhanced by adding more gates to the PRISM modules library which include the Priority-AND, INHIBIT, k/N gates and to incorporate lumping of states as in ~\cite{yevkin2015efficient}.
	
	\begin{acks}
		{
			The author's would also like to thank Carlos E. Budde and Enno Ruijters for their useful discussion and suggestions.
			This work has been funded by the {AMBI project} under Grant No.: {324432}, by the Alan Turing Institute, UK, post-doctoral research grant from {Fonds de Recherche du Quebec - Nature et Technologies (FRQNT)} and Malta's ENDEAVOUR Scholarships Scheme.}
	\end{acks}

	\bibliographystyle{ACM-Reference-Format}
	\bibliography{Bib_rep}

\end{document}